\documentclass[12pt]{iopart}
\usepackage{iopams,graphicx}  
\begin{document}

\title{Inherent Stochastic Linearization of Random Laser Modes}

\author{Jonathan Andreasen$^{1,2*}$ and Hui Cao$^{1,3}$}

\address{$^1$Department of Applied Physics, Yale University, New Haven, CT 06520\\
$^2$Laboratoire de Physique de la Mati\`ere Condens\'ee, CNRS UMR 6622, \\
Universit\'e de Nice-Sophia Antipolis, Parc Valrose, 06108, Nice Cedex 02, France\\
$^3$Department of Physics, Yale University, New Haven, CT 06520}

\ead{jonathan.andreasen@unice.fr}

\begin{abstract}
  Weakly scattering random lasers exhibit lasing modes that spatially overlap and can interact strongly via gain saturation.
  Consequently, lasing in high-threshold modes may be suppressed by strong low-threshold lasing modes.
  We numerically examine the effect of inherent noise on this strong nonlinear phenomenon.
  Noise generates emission below the lasing threshold and 
  restrains the dramatic nonlinear behavior above threshold. 
  The result is a linearization of random laser modes and is possible
  when noise overcomes spatial hole burning.
  Results suggest that control over the noise properties of the gain medium may 
  facilitate or inhibit certain modes to lase in the multimode regime.
\end{abstract}


\section{Introduction}

Contrary to conventional lasers, random lasers have no cavity like a Fabry-P{\'e}rot resonator \cite{caoLRM}.
Instead, they are made of a multiply scattering medium such as a semiconductor powder \cite{markushev,cao99} or a 
suspension of scattering particles in dye solution \cite{cao00}, which is excited by an external pump to introduce gain.
Multiple scattering of light in the random medium provides optical feedback and
lasing modes are built on the quasimodes of the passive random system.
A recent review devoted to the first lasing mode at threshold \cite{review}
shows how the relation between lasing modes and quasimodes depends on the openness of the system.
With strong confinement of light, as in the localization regime, lasing modes have a nearly one-to-one
correspondence with the localized modes of the passive system.
In diffusive systems, quasimodes exhibit a large amount of spatial and spectral overlap but
maintain a strong correspondence with the lowest threshold lasing modes.
In systems which are more open, such as those in the quasi-ballistic regime, the correspondence significantly degrades.
This is largely due to the intense pumping required to overcome high loss from the openness,
which introduces a modification of the refractive index distribution.

Far above the lasing threshold, in the case of multimode lasing, it was found that the correspondence between lasing modes and quasimodes
begins to degrade in the diffusive regime \cite{tureciSci,tureciNL09}.
Mode competition occurs due to gain saturation. 
With limited gain available, spatial hole burning takes place 
where the field intensity is large.
Thus, random lasing thresholds may increase in the multimode regime and lasing in some modes may be completely suppressed.
In other words, ``dead'' regions are produced in the spatial profile of gain
caused by the low-threshold lasing modes which proves detrimental for lasing in other higher-threshold modes.
Illustrations of such strong nonlinear effects were made by
taking into account the openness of the system and the nonlinearity to all orders via
steady-state \textit{ab initio} laser theory \cite{tureciPRA}.
Nonlinear effects above threshold have also been studied in the time domain with full-wave simulations incorporating four-level atomic media \cite{andreasenNLE}. 
However, the effects of intrinsic noise, which cause dynamic changes to the atomic population and polarization \cite{andreasen09jlt},
on nonlinear processes has not yet been taken into account.

In this paper, a frequency-dependent linear gain model is first employed to examine lasing modes without the effects of gain saturation.
Gain saturation is then incorporated via full-wave Maxwell-Bloch simulations.
Comparison of the two methods determines the nonlinear effects of gain saturation.
Intrinsic noise in weakly scattering random lasers has been shown \cite{andreasenrln} to alter lasing thresholds and 
introduce peaks in the emission spectra which were absent from the spectra without noise.
Thus, we examine how noise modifies the dramatic nonlinear effects introduced by gain saturation.
The population inversion is found to be significantly affected by noise in the multimode lasing regime not far from the lasing threshold.
Spatial hole burning can be overcome to excite and amplify additional modes.
Finally, modal amplitudes are found to be linearized across the lasing threshold and in some cases, mode suppression can be mitigated.

This paper is organized as follows.
In section \ref{sec:nummeth}, information concerning the numerical methods employed in this paper is given.
In section \ref{sec:active}, nonlinear effects above the first threshold for lasing are studied without noise.
In section \ref{sec:noise}, the inherent noise of optical systems is taken into account.
Spatial properties of the gain are examined in section \ref{sec:spatial}.
Finally, inherent stochastic linearization of random laser modes is discussed in section \ref{sec:stoch}.
Final conclusions are presented in section \ref{sec:conclusion}.

\section{Numerical methods\label{sec:nummeth}}

\subsection{Random Structure}

The one-dimensional random systems considered are composed of $41$ layers.
Dielectric material with index of refraction $n_1=2$ separated by air gaps
($n_2=1$) results in a spatially modulated index of refraction $n(x)$.
Outside the random medium $n_0 = 1$.
The average thicknesses are $\left< d_1 \right> = 100$ nm and $\left<d_2\right> = 200$ nm
giving a total average length of $\left<L\right> = 6100$ nm.
In the wavelength range of interest (400 nm -- 800 nm),
the localization length $\xi$ ranges from 850 nm to 1500 nm.
$\xi$ was calculated from the dependence of ensemble-averaged transmittance $T$ 
on the system lengths $L$ as $\xi^{-1} = -d\left<\ln T\right>/dL$.
The Thouless number $g$, which reveals the amount of spectral overlap of resonances of these
random systems, is given by the ratio of the average resonance linewidth to the average
frequency spacing $g=\left<k_i\right>/\left<dk\right>$.
The linewidth is estimated via the spectral correlation function $G(\Delta k)$ of the transmission $T(k)$
\begin{equation}
  G(\Delta k) = \left<T(k)T(k+\Delta k)\right>/\left<T^2(k)\right>.
\end{equation}
The width of $G(\Delta k)$ estimates $\left<k_i\right>$.
The resonance frequencies of the passive system (found via the transfer matrix method described below)
are used to estimate $\left<dk\right>$.
This results in $g=0.18$, meaning the resonances are well separated.

In such strongly scattering systems, spectrally separated modes generally exhibit less spatial overlap
and thus, less interaction \cite{caoPRB03,gePRA10}.
However, we shall show that lasing modes can still interact strongly through the gain medium 
whose homogeneously broadened spectrum covers multiple resonances.

\subsection{Frequency-dependent Linear Gain Model}

The transfer matrix (TM) method developed in \cite{andreasennu} is used to simulate
lasing modes at threshold with linear gain.
Gain is linear in that it does not depend on the electromagnetic field intensity.
Thus, gain saturation is not included and consequently, mode interactions via spatial hole burning are neglected.
Solutions are only valid at or below threshold \cite{souk99},
not above it where gain saturation is needed to reach a steady state.

The lasing solutions must satisfy the time-independent wave equation
with a complex frequency-dependent dielectric function
\begin{equation}
  \epsilon(x,k) = \epsilon_r(x) + \chi_g(x,k),
\end{equation}
where a real wavenumber $k = 2\pi/\lambda$ describes the light frequency in vacuum, 
and $\epsilon_r(x)=n^2(x)$ is the dielectric function of the passive
background material.
The frequency dependence of $\epsilon_r$ is ignored.
$\chi_g(x,k)$, corresponding to the susceptibility of the atomic material, is given by
\begin{equation}
  \chi_g(x,k)=\frac{-A_eN_A(x)}{k_a^2-k^2
    -ik\Delta k_a},
  \label{eq:chia1}
\end{equation}
where $A_e$ is a material-dependent constant, $N_A(x)$ is the spatially dependent
density of atoms, $k_a$ is the atomic transition frequency, and
$\Delta k_a$ is the spectral width of the atomic resonance.
Real quantum transitions may be considered \cite{siegbook} to induce a response
proportional to the population difference density $\Delta N_A$.
$N_A(x)$ is thus replaced in equation (\ref{eq:chia1}) by $\Delta N_A(x)=N_2(x)-N_1(x)$, the difference in population
between the upper and lower energy levels (i.e., population inversion). 
$\epsilon(x,k)$ therefore includes absorption [$\Delta N_A(x)<0$] or gain [$\Delta N_A(x)>0$].
The complex, frequency-dependent index of refraction used in the TM method is calculated as 
\begin{equation}
  \tilde{n}(x,k) = \sqrt{\epsilon_r(x) + \chi_g(x,k)}.
\end{equation}

The atomic transition frequency is set to $k_a=10.5$ $\mu$m$^{-1}$, the corresponding wavelength
$\lambda_a=600$ nm.
The width of the gain spectrum is chosen such that it spans ten resonances of the passive system,
giving $\Delta k_a=3.7$ $\mu$m$^{-1}$ and in wavelength-space, $\Delta\lambda_a = 200$ nm. 
Propagation of the electric field through the structure is calculated via
the $2\times 2$ matrix $M$.
Boundary conditions with only emission out of the system require $M_{22}=0$.
We consider a spatially uniform population inversion ($\Delta N_A(x)\rightarrow \Delta N_A$) to avoid
additional light scattering caused by the spatial inhomogeneity of gain (imaginary part of $\tilde{n}(x,k)$).
Although it does not correspond to common experimental situations where gain atoms are
incorporated only in the dielectric layers, it is possible to have gain atoms in the
gas phase distributed in the air gaps.
Lasing frequencies and thresholds are determined by finding the values of $k$ and $\Delta N_A$, respectively, that satisfy $M_{22}=0$.

\subsection{Maxwell-Bloch Equations\label{sec:ran}}

This numerical method is based on the finite-difference time-domain (FDTD) formulation we
developed to study the effects of noise on light-atom interaction in complex systems without prior
knowledge of resonances \cite{andreasen08, andreasen09jlt}.
Two-level atoms are uniformly distributed over the entire random system to avoid
additional light scattering caused by the spatial inhomogeneity of gain.
The two-level model of atoms is a simplified approach that can be applied to actual lasers
based on three-level atoms such as Ruby and Erbium lasers, as the population in the third
level is negligibly small \cite{valcarcelRMF06}.

The atomic transition frequency is set to $k_a=10.5$ $\mu$m$^{-1}$, 
the corresponding wavelength $\lambda_a=600$ nm.
The lifetime of atoms in the excited state $T_1$ and the dephasing time $T_2$ are included in the Bloch equations.
The width of the gain spectrum is given by $\Delta k_a=(1/T_1+2/T_2)/c$.
We set $T_1=1.0$ ps.
The value of $T_2=1.8$ fs is chosen such that the gain spectrum spans ten resonances of the passive system.
We also include incoherent pumping of atoms from level 1 to level 2.
The rate of atoms being pumped is proportional to the population of atoms in level 1 [$\rho_{11}(x)$]
and the proportionality coefficient $P_r$ is called the pumping coefficient.
In the steady state, a spatially-dependent population inversion $\rho_3(x)=\rho_{22}(x) - \rho_{11}(x)$ 
emerges and is within the interval [-1, 1].
This quantity is spatially averaged and represents the pump level.
This number can be compared with the threshold population inversion $\Delta N_A$ found via the TM method.
The spatial properties of $\rho_3(x)$ are also examined.
These Maxwell-Bloch (MB) simulations solve for the atomic population inversion $\rho_{3}(x)$ 
and atomic polarization $\rho_1(x)=\rho_{12}(x) + \rho_{21}(x)$ and $\rho_2(x)=i[\rho_{12}(x) - \rho_{21}(x)]$.

To introduce noise to the MB equations, we use the stochastic $c$-number equations that are
derived from the quantum Langevin equations in the many-atom limit \cite{drum91}.
Based on the fluctuation-dissipation theorem, noise accompanies decay of the light field and atomic dissipation.
The amplitude of classical noise accompanying the field decay is proportional to the square root of
the thermal photon number.
At room temperature the number of thermal photons at visible frequencies is negligible.
Thus the noise related to field decay is ignored here.
At higher temperatures or longer wavelengths, this noise becomes significant.
The thermal noise and its temporal coherence can be incorporated into the FDTD algorithm following the approach we developed 
in our previous work \cite{andreasen08}.

We consider noise associated with three dissipation mechanisms for atoms (described in detail in \cite{andreasen09jlt})
(i) dephasing events,
(ii) excited state decay,
(iii) incoherent pumping (from ground state to excited state).
The stochastic MB simulations solve for the atomic population inversion
$\rho_{3}(x)$ and atomic polarization $\rho_1(x)$ and $\rho_2(x)$.
With $T_2 \ll T_1$, we neglect the influence of population fluctuations on the polarization
because it is orders of magnitude smaller than noise due to dephasing.
All calculations here are done in the regime $\rho_{22} \gtrsim \rho_{11}$,
where stochastic terms in the density matrix evolution of the \emph{macroscopic}
system successfully mimic spontaneous emission \cite{andreasen09jlt,sukharev11}. 

\section{Nonlinear Effects Above the First Lasing Threshold\label{sec:active}}

\subsection{Effects of Gain Saturation\label{ssc:gainsat}}

\begin{figure}
  \begin{center}
    \includegraphics[width=13cm]{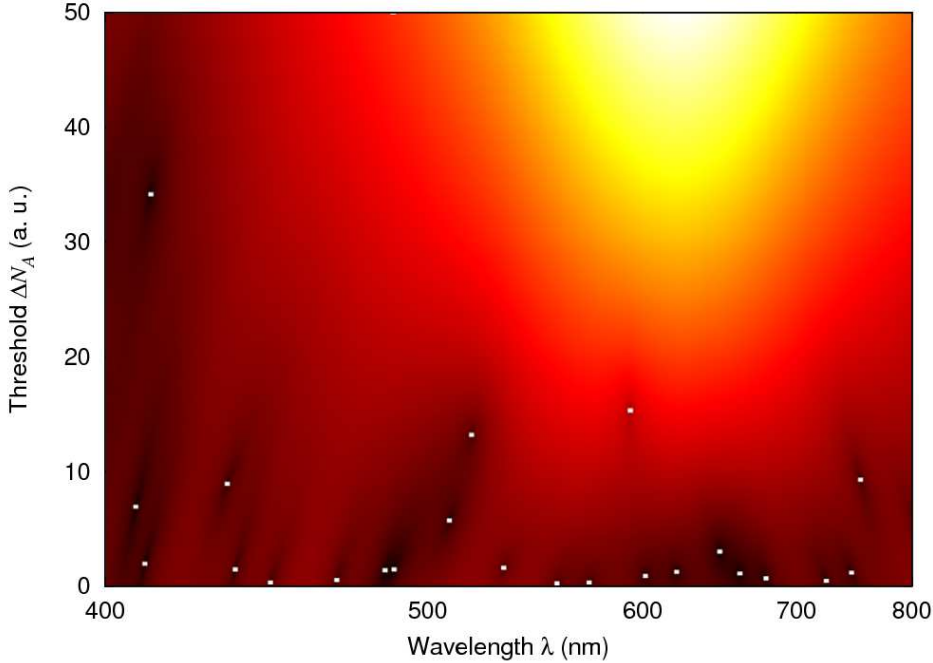}
    \caption{\label{fig:fig1} 
      Lasing solutions with frequency-dependent linear gain (TM method) shown via a
      map of $M_{22}(\lambda$, $\Delta N_A)$, where dark regions indicate values near zero. 
      Gain saturation is neglected.
      Lasing modes are marked by white squares.      
      23 modes are found in the wavelength range with a wide distribution of thresholds.
      The bright region near $\lambda_a=600$ nm indicates large values of $M_{22}$,
      meaning large-threshold modes may not exist near the gain center wavelength.
    }
  \end{center}
\end{figure}

Gain saturation is first neglected in the frequency-dependent TM calculation and the lasing thresholds of the random system found.
Figure \ref{fig:fig1} maps the wavelengths and thresholds ($\lambda$, $\Delta N_A$) of lasing modes.
Many possible lasing modes exist due to the width of the gain spectrum.
In this strongly scattering system, the intensity distributions of modes are distributed throughout the 
structure, but still fully contained inside the structure ($\xi < L$).
Such systems possess a wide distribution of decay rates \cite{mirlinpr00,terraneoEPJB00,pinheiroPRE04}.
This translates into a wide distribution of lasing thresholds \cite{patraPRE03,apalkovPRB05}.
The effect is seen clearly in figure \ref{fig:fig1} where threshold values extend over an order of magnitude. 
Near the gain center wavelength, amplification is large and thus, lasing modes have smaller thresholds.
Large-threshold modes are not observed in this region as indicated by the bright region where $M_{22}$ is far from zero.
We focus here on the small-threshold modes (small $\Delta N_A$). 

\begin{figure}
  \begin{center}
    \includegraphics[width=6cm,angle=270]{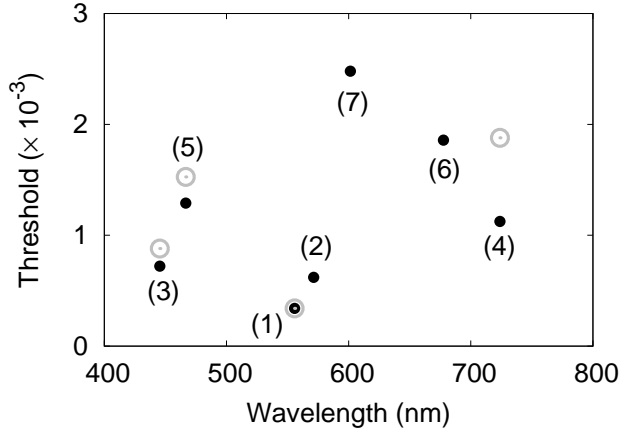}
    \caption{\label{fig:fig2}
      Threshold population inversion of the first 7 lasing modes without (TM--filled circles)
      and with (MB--open circles) gain saturation included.
      TM thresholds are normalized relative to the first MB threshold.
      Modes are labeled in order of increasing threshold without gain saturation.
    }
  \end{center}
\end{figure}

Figure \ref{fig:fig2} reveals the wavelengths and thresholds of the first 7 lasing modes.
Both vary stochastically due to the randomness of the structure. 
In general, modes closer to $\lambda_a=600$ nm have smaller thresholds. 
However, some modes farther from the gain center (e.g., mode 4) are associated with quasimodes which have smaller decay rates
and thus, have smaller thresholds.

Effects of gain saturation are examined here through MB simulations (\emph{without noise}).
Emission spectra are found through fast Fourier transformations of the steady-state output field.
A Welch window is used to keep large-amplitude peaks from overlapping and thereby masking small-amplitude peaks in the spectra.
Lasing thresholds are determined to be at the lowest pump level at which a peak appears in the emission spectrum.
The steady-state spatiotemporally averaged population inversion is compared to the
threshold population inversion found via the TM method.
The first lasing threshold should be the same for both the TM and MB calculations since, at this point,
there is no gain saturation nor mode competition.
Thus, the first TM threshold is normalized to the first MB threshold.
All other TM thresholds are scaled by the same ratio and
the results are shown in figure \ref{fig:fig2}. 

\begin{figure}
  \begin{center}
    \includegraphics[width=5.3cm,angle=270]{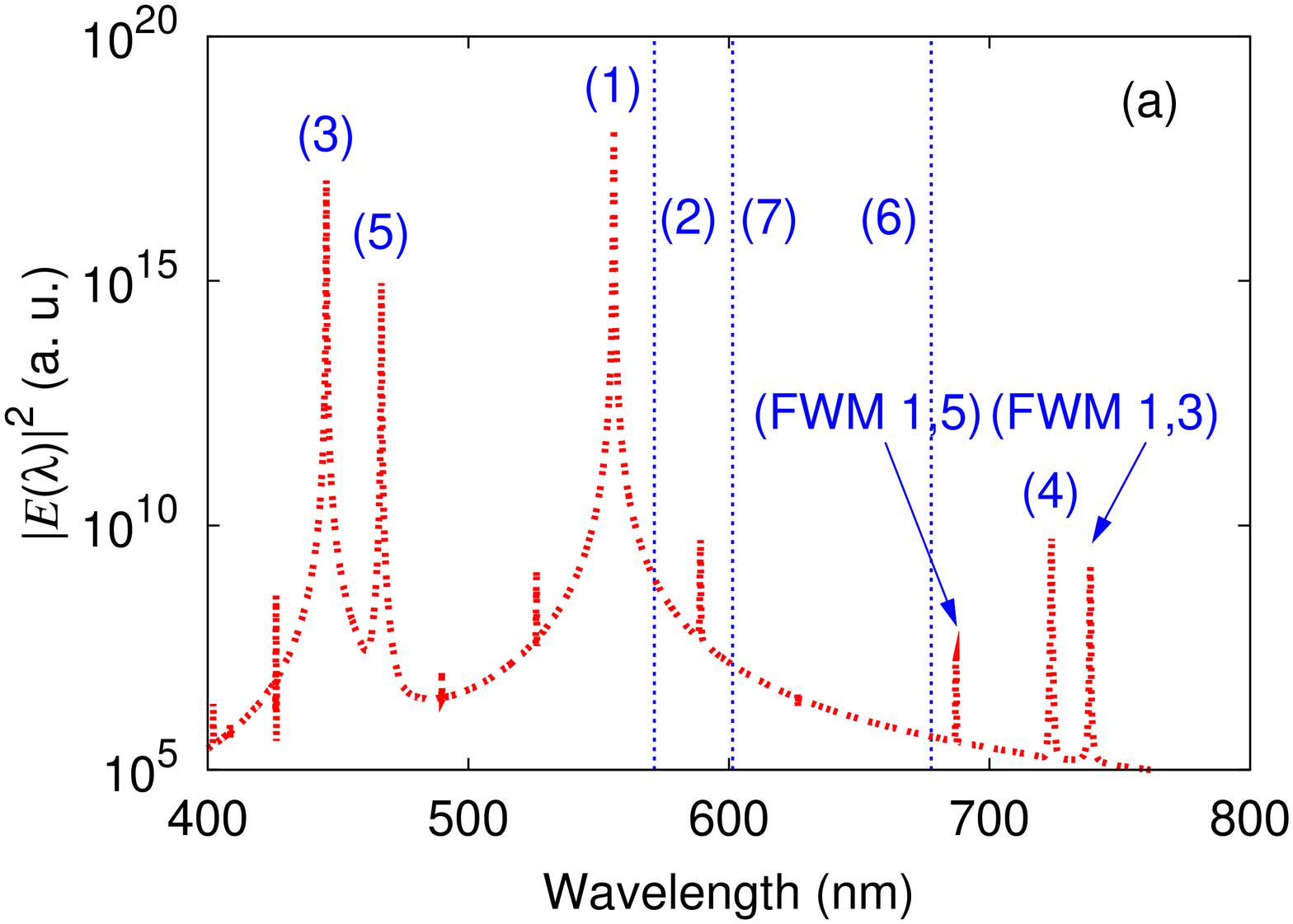}
    \includegraphics[width=5.3cm,angle=270]{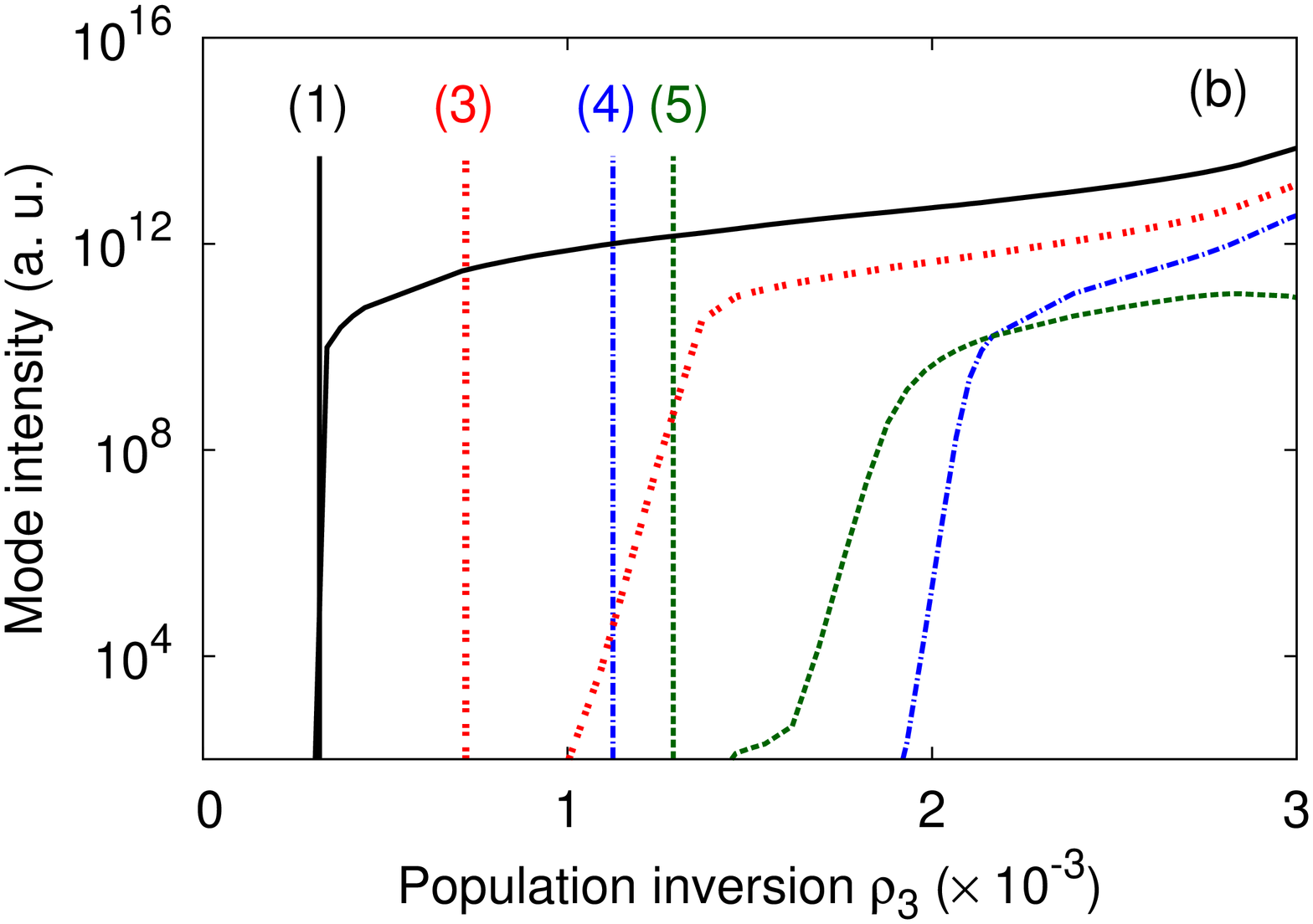}
    \caption{\label{fig:fig3}
      (a)
      Emission spectrum from MB simulations with gain saturation for $\left<\rho_3(x,t)\right>=2\times 10^{-3}$.
      Lasing modes are enumerated in figure \ref{fig:fig2}.
      The four lasing peaks are labeled (1, 3, 4, 5) and the wavelengths of the three suppressed lasing modes (2, 6, 7) 
      are indicated by vertical blue lines.
      Various other peaks appear in the emission spectrum due to nonlinear wave-mixing.
      Two four-wave mixing peaks (involving modes 1, 3, and 5) are labeled as examples.
      (b)
      Intensity of random lasing modes with increasing $\left<\rho_3(x,t)\right>$.
      Vertical lines indicate the corresponding TM thresholds.
      Modes 4 and 5 switch order with gain saturation included.
    }
  \end{center}
\end{figure}

MB thresholds of modes 3, 4, and 5 are larger than their TM counterparts.
This is expected \cite{tureciSci,tureciNL09,andreasenNLE} since spatial hole burning caused by the first mode reduces the gain available for larger-threshold modes, 
thereby increasing their thresholds.
Three modes, 2, 6, and 7, are missing from the MB simulations.
The emission spectrum near the lasing threshold of mode 4 
is shown in figure \ref{fig:fig3}(a) to verify this behavior.
It is clear that mode 2 is suppressed in this case, i.e., it is not lasing.
Higher pump levels were checked but modes 6 and 7 were not found.
Their behavior shall later be discussed in more detail.

Figure \ref{fig:fig3}(b) reveals the intensity of lasing modes as a function of the pump level above threshold.
A sharp increase of mode intensity is seen near the lasing threshold,
which is larger than the corresponding TM threshold indicated by a vertical line.
Within a small range of pump levels (approximately $1.5\times 10^{-3} < \left<\rho_3(x,t)\right> < 2 \times 10^{-3}$),
mode 4 is suppressed while mode 5 lases.
Eventually mode 4 reaches its lasing threshold and quickly overtakes mode 5 in intensity.
In general, stronger scattering systems have less mode overlap and therefore weaker competition effects \cite{gePRA10}.
However, the relatively large number of lasing modes (due to a wide gain spectrum)
encourages mode interaction and stimulates mode suppression.

\subsection{Nonlinear Wave-Mixing\label{sec:nonlinear}}

Nonlinear wave-mixing in random media is well known \cite{millerPR64,deweyAPL75,baudrierN04,skipetrovN04}
and occurs regularly due to random quasi-phase-matching.
However, such effects have only been observed recently in random lasers through numerical simulations
with four-level gain atoms \cite{andreasenNLE}. 
Here, with two-level gain atoms, a higher pump level results in 
four-wave mixing (FWM) involving modes 1 and 3 seen in figure \ref{fig:fig3}(a), 
with a peak at $(2\lambda_1^{-1} - \lambda_3^{-1})^{-1}\approx 740$ nm.
Another peak, with mixing involving modes 1 and 5, is seen at $(2\lambda_1^{-1} - \lambda_5^{-1})^{-1}\approx 690$ nm.
Many such peaks exist and other nonlinear processes, such as third-harmonic generation,
occur simultaneously but at much shorter wavelengths.

The amplitude of FWM peaks is orders of magnitude smaller than the lasing peaks with which they are associated.
The FWM peaks may not generally influence steady-state lasing properties due to their small amplitudes.
However, they can be comparable in amplitude to higher-threshold lasing modes
[e.g., lasing peak 4 and FWM peak (1,3)] in figure \ref{fig:fig3}(a)]. 

\begin{figure}
  \begin{center}
    \includegraphics[width=6cm,angle=270]{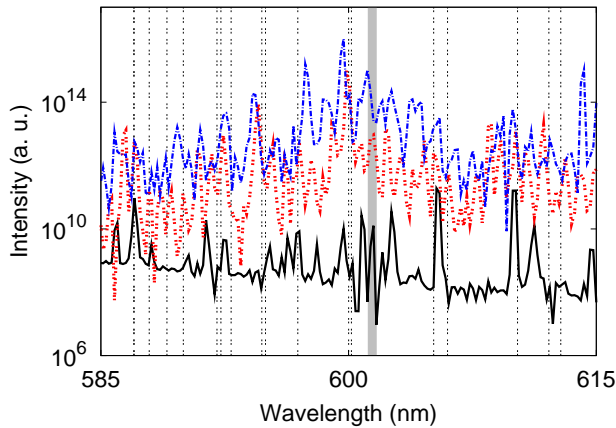}
    \caption{\label{fig:fig4} 
      Emission spectra from MB simulations \emph{without} noise for 
      $\left<\rho_3(x,t)\right>=$ 
      (black solid line) $3.84\times 10^{-3}$,
      (red dotted line) $5.89\times 10^{-3}$, and
      (blue dashed line) $6.84\times 10^{-3}$.
      The spectrum is focused around $\lambda_7=601.4$ nm (thick vertical gray line).
      Though noise is not included in these simulations, the spectrum appears noisy
      due to nonlinear wave-mixing (FWM wavelengths marked by vertical black dotted lines).
      A broad distribution of intensity appears around $\lambda_7$ for high pump levels.
    }
  \end{center}
\end{figure}

Another example showing the influence of FWM peaks is shown in figure \ref{fig:fig4},
where mode 7 is examined.
The lasing threshold for mode 7 predicted by the TM method is $\left<\rho_3(x,t)\right>=2.5\times 10^{-3}$.
Above this pump level, a multitude of FWM peaks are generated at wavelengths close to mode 7, which obfuscates the character of the mode.
The signal integrated around the mode 7 wavelength $\lambda_7$ is orders of magnitude smaller than the signal of known 
lasing modes at the same pump level.
However, it appears that the fields generated by FWM are somewhat trapped by the resonance at $\lambda_7$ resulting in a 
broad peak around $\lambda_7$ in figure \ref{fig:fig4}.
The center of the broad peak does not exactly coincide with $\lambda_7$,
but this could be attributed to frequency pulling caused by the large amount of gain.
Similar behavior occurs for modes 2 and 6.
We conclude that these modes do not lase without noise.
Fields due to FWM can be trapped at these resonances, but
a thorough examination of this effect on lasing of large threshold modes is beyond the scope of this paper.

\section{Impact of Noise\label{sec:noise}}

\subsection{Spectral Behavior\label{ssc:spectbeh}}

\begin{figure}
  \begin{center}
    \includegraphics[width=6cm,angle=270]{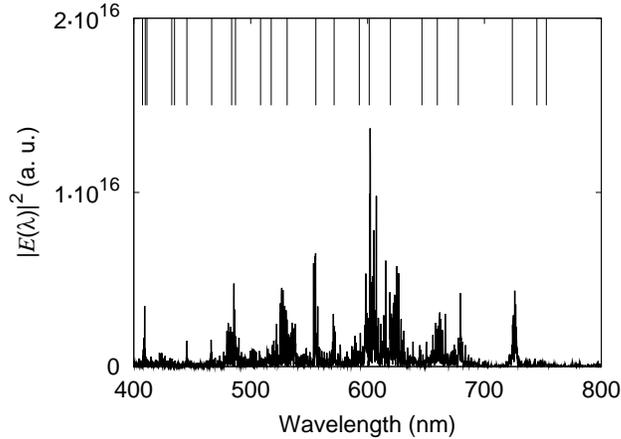}
    \caption{\label{fig:fig5}
      Emission spectra from MB simulation with noise at the transparency point ($\left<\rho_3(x,t)\right>=0$).
      (top) Vertical lines mark the resonance wavelengths found via the TM method.
    }
  \end{center}
\end{figure}

We first introduce the emission spectra for the pump level at the transparency point then investigate higher pump levels.
Without noise, at a pumping coefficient of $P_r=1.00$, the system becomes transparent
($\rho_3=\rho_{22}-\rho_{11}=0$ at the steady state).
Since there is no net gain, the initial seed pulse dies away, and there is no signal at the steady state.
Noise slightly reduces the excited state population \cite{milonnil},
thus the system is just below the transparency point for $P_r=1.00$.
The pumping coefficient $P_r$ can be adjusted so that the steady-state spatiotemporally 
averaged population inversion is zero ($\left<\rho_3(x,t)\right>=0$).
Figure \ref{fig:fig5} shows the steady-state emission spectrum with noise $|E(\lambda)|^2$ at this point
when there is no net gain nor absorption.
The spectrum is broad and centered at the atomic transition wavelength $\lambda_a=600$ nm.

Spectral modulation of emission intensity is evident in figure \ref{fig:fig5} though there is no net gain.
Without amplification, the system cannot support lasing.
However, due to strong scattering, the dwell time of light at resonant frequencies is longer than at nonresonant frequencies
so the field builds up in the system.
Peaks due to this buildup are visible because the resonance peaks are spectrally separated ($g=0.18$).
Thus, modes are visible as peaks in the emission spectrum without gain.

\begin{figure}
  \begin{center}
    \includegraphics[width=5.3cm,angle=270]{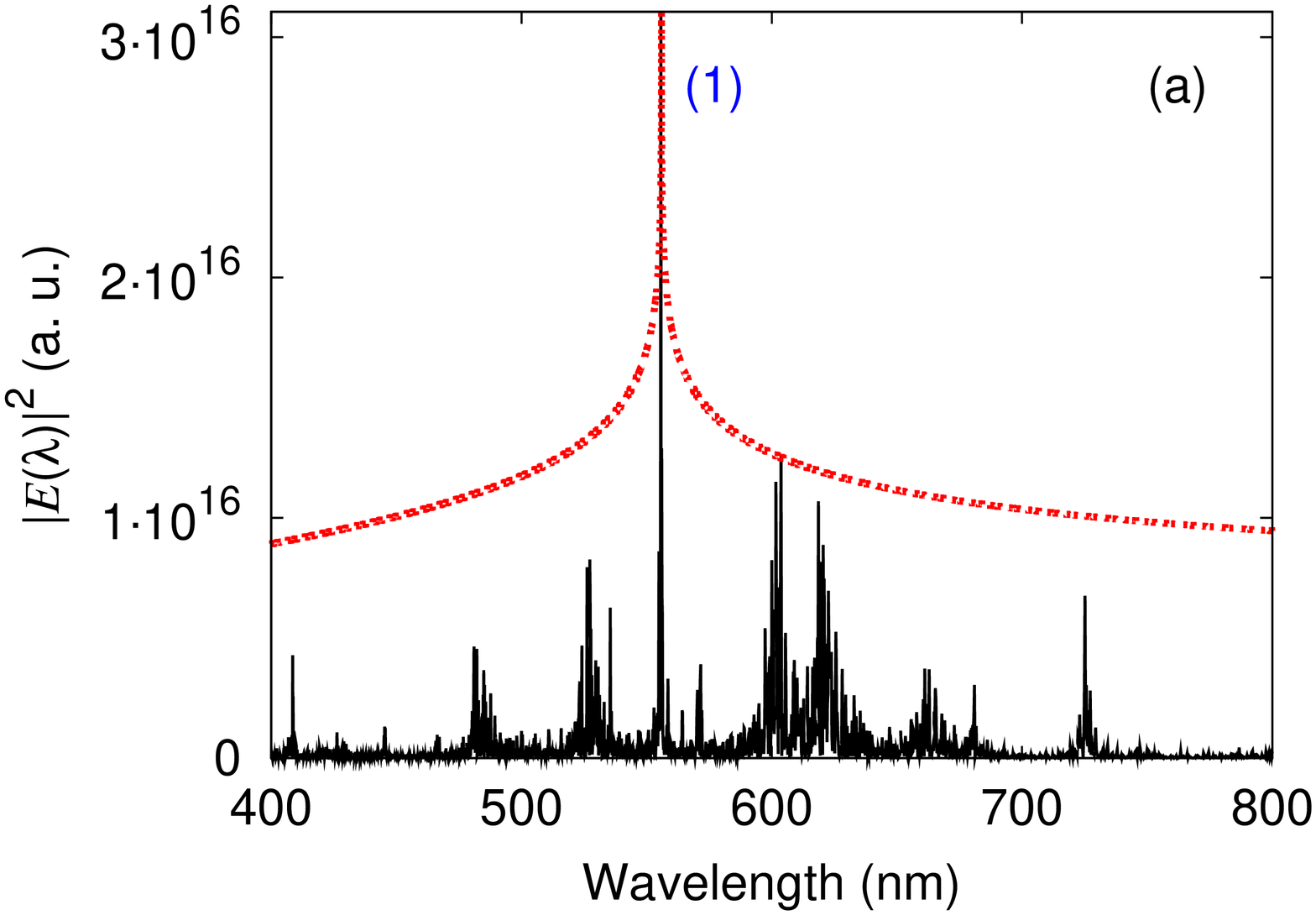}
    \includegraphics[width=5.3cm,angle=270]{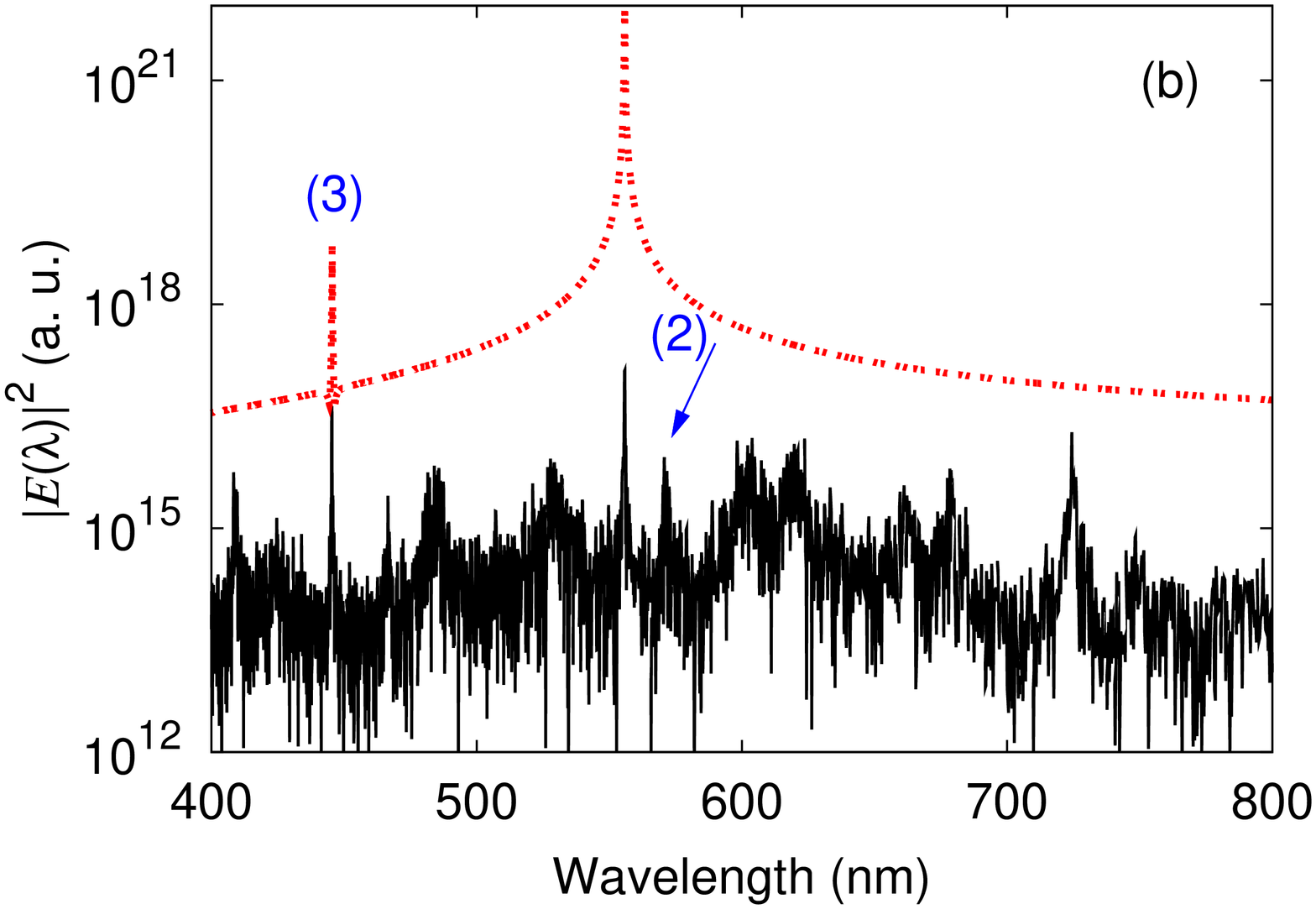}\\
    \includegraphics[width=5.3cm,angle=270]{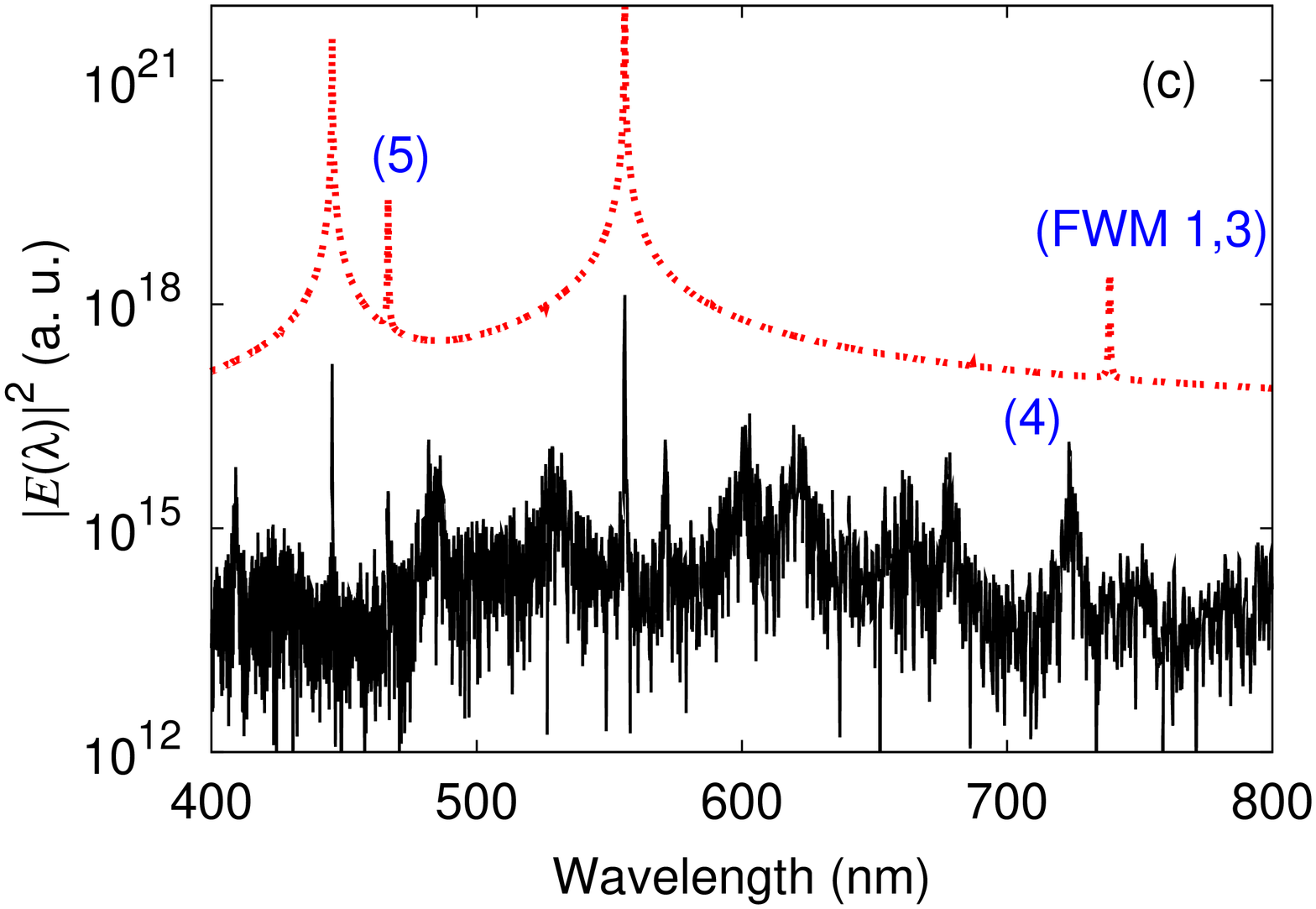}
    \includegraphics[width=5.3cm,angle=270]{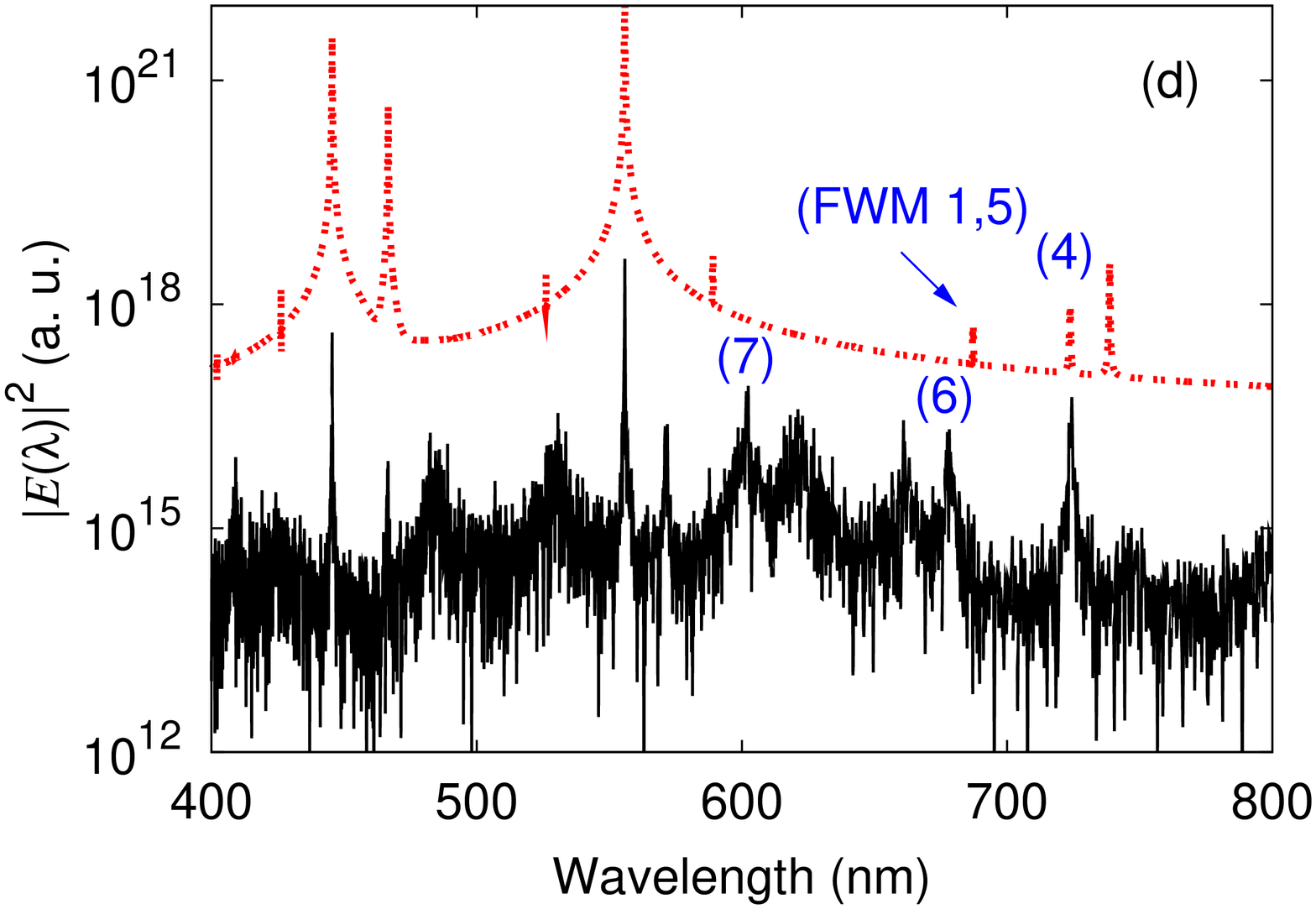}\\
    \caption{\label{fig:fig6}
      Emission spectra without noise (red) and
      with noise (black) for 
      $\left<\rho_3(x,t)\right>=$ 
      (a) $0.30\times 10^{-3}$,
      (b) $1.1\times 10^{-3}$,
      (c) $1.8\times 10^{-3}$,
      (d) $1.9\times 10^{-3}$.
      Modes [enumerated in figure \ref{fig:fig2}] are marked in sequence
      along with along with relevant peaks due to four-wave mixing.
      Spectra without noise are vertically offset and normalized for clarity.
    }
  \end{center}
\end{figure}

Introducing amplification ($\rho_3 > 0$) allows the first lasing threshold to be reached without noise
at $\left<\rho_3(x,t)\right>=0.3\times 10^{-3}$ in figure \ref{fig:fig6}(a).
The single lasing peak matches the lasing mode 1 wavelength found in the absence of gain saturation. 
A narrow spectral peak also appears at the same wavelength with noise.
Due to the smooth transition from amplified spontaneous emission (ASE) to lasing \cite{andreasenrln}, determining lasing thresholds 
with noise is nontrivial and shall be discussed later.
Meanwhile, we observe that most of the resonance peaks exist in the spectrum with noise (though some may be buried)
and become narrower by light amplification. 

With gain saturation included, the second mode to reach its lasing threshold in the absence of noise [figure \ref{fig:fig6}(b)]
is mode 3 (enumerated in figure \ref{fig:fig2}). 
Gain saturation evidently causes mode 2 to be suppressed. 
With noise, however, both modes 2 and 3 are seen in the emission spectrum.
The mode 2 peak has a smaller amplitude and a larger linewidth than mode 3.

Mode 5 is next to reach the lasing threshold without noise in figure \ref{fig:fig6}(c)
meaning mode 4 is suppressed.
Again, mode 4 is observed in the emission spectrum with noise but is slightly stronger than mode 5 in this case.
Even though mode 4 is farther from the gain center wavelength, its amplitude is comparable to that of mode 5.
The linewidth of mode 4 is also slightly narrower than that of mode 5.

Without noise, mode 4 begins lasing at a higher pump level as shown in figure \ref{fig:fig6}(d).
A corresponding peak is seen easily in the spectrum with noise. 
Though much higher pump levels 
were checked, modes 2, 6, and 7 are never seen clearly in the emission spectrum without noise.
The peak in the spectrum without noise [in figure \ref{fig:fig6}(d)], whose frequency is close to that of mode 6,
is in fact a FWM peak involving modes 1 and 5, i.e., $(2\lambda_1^{-1} - \lambda_5^{-1})^{-1}$.
All three modes (2, 6, 7), however, clearly exist in the spectra with noise.

Note that although modes appear as peaks in the emission spectra with noise, 
FWM  peaks are not clearly observed.
It is unclear if the FWM peaks are merely hidden in the noise background or if FWM is suppressed by noise.
One possibility is that noise continually randomizes the phases of modes making even random quasi-phase-matching
difficult to achieve.

\subsection{Lasing Threshold\label{sec:noisethresh}}

\begin{figure}
  \begin{center}
    \includegraphics[width=3.4cm,angle=270]{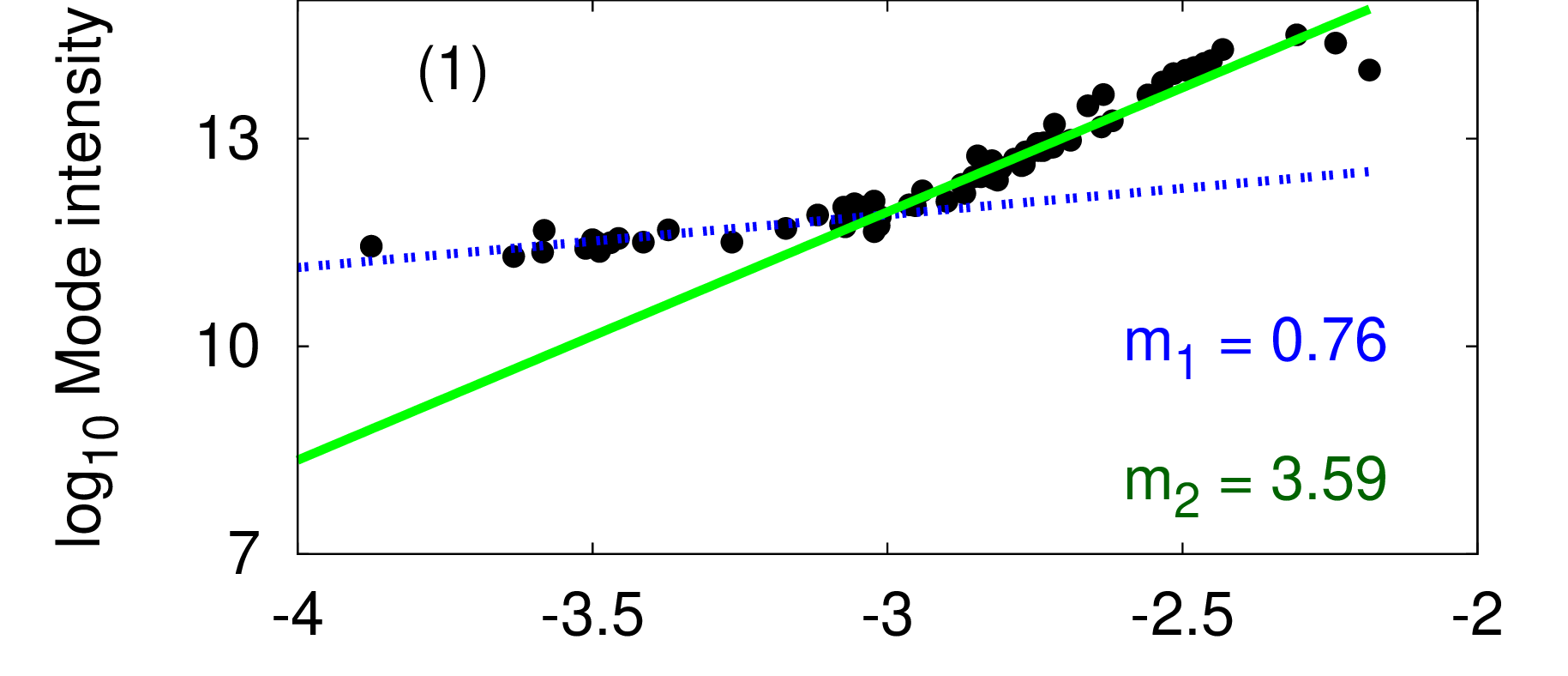}
    \includegraphics[width=3.4cm,angle=270]{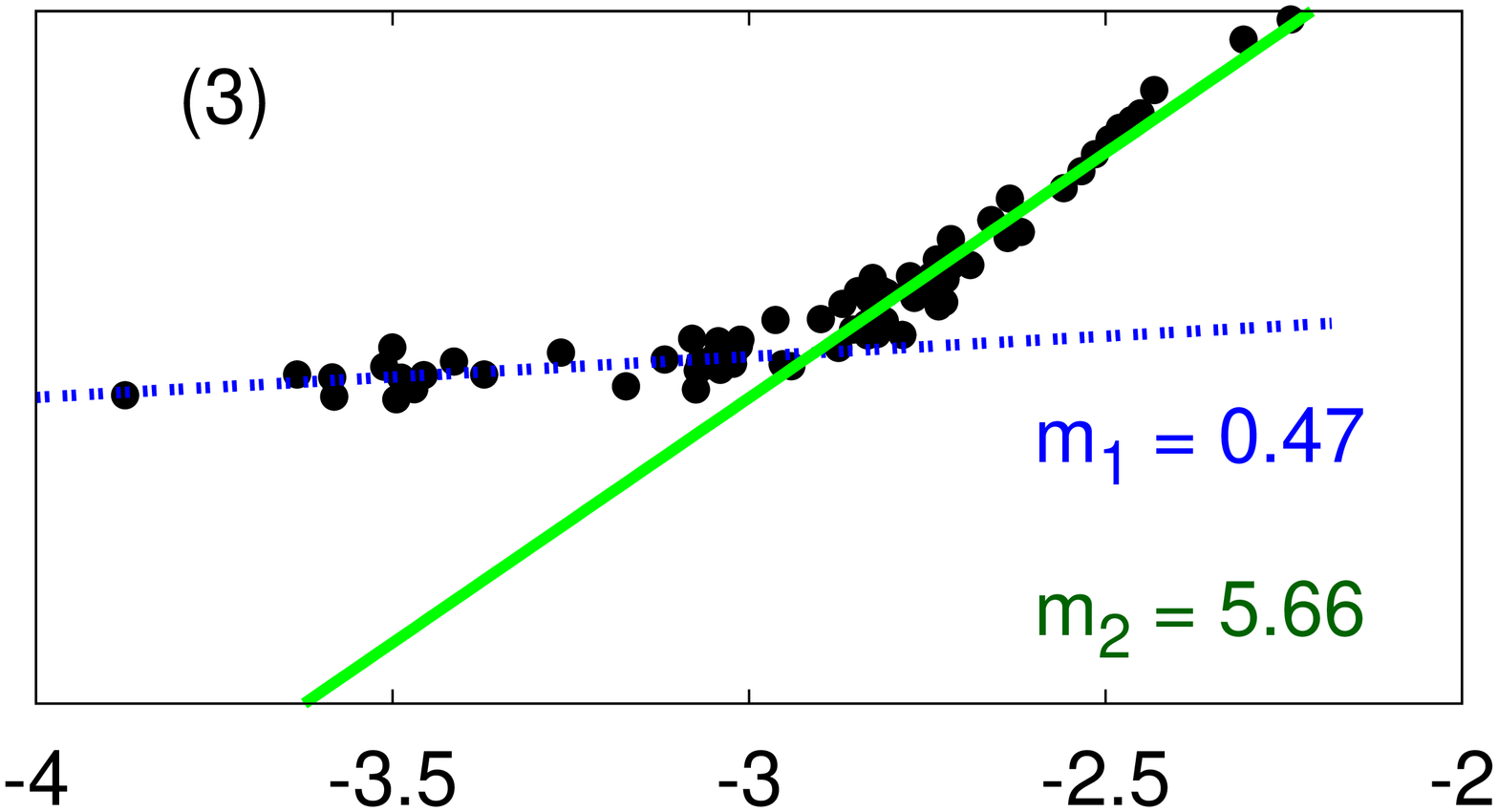}\\
    \includegraphics[width=4.5cm,angle=270]{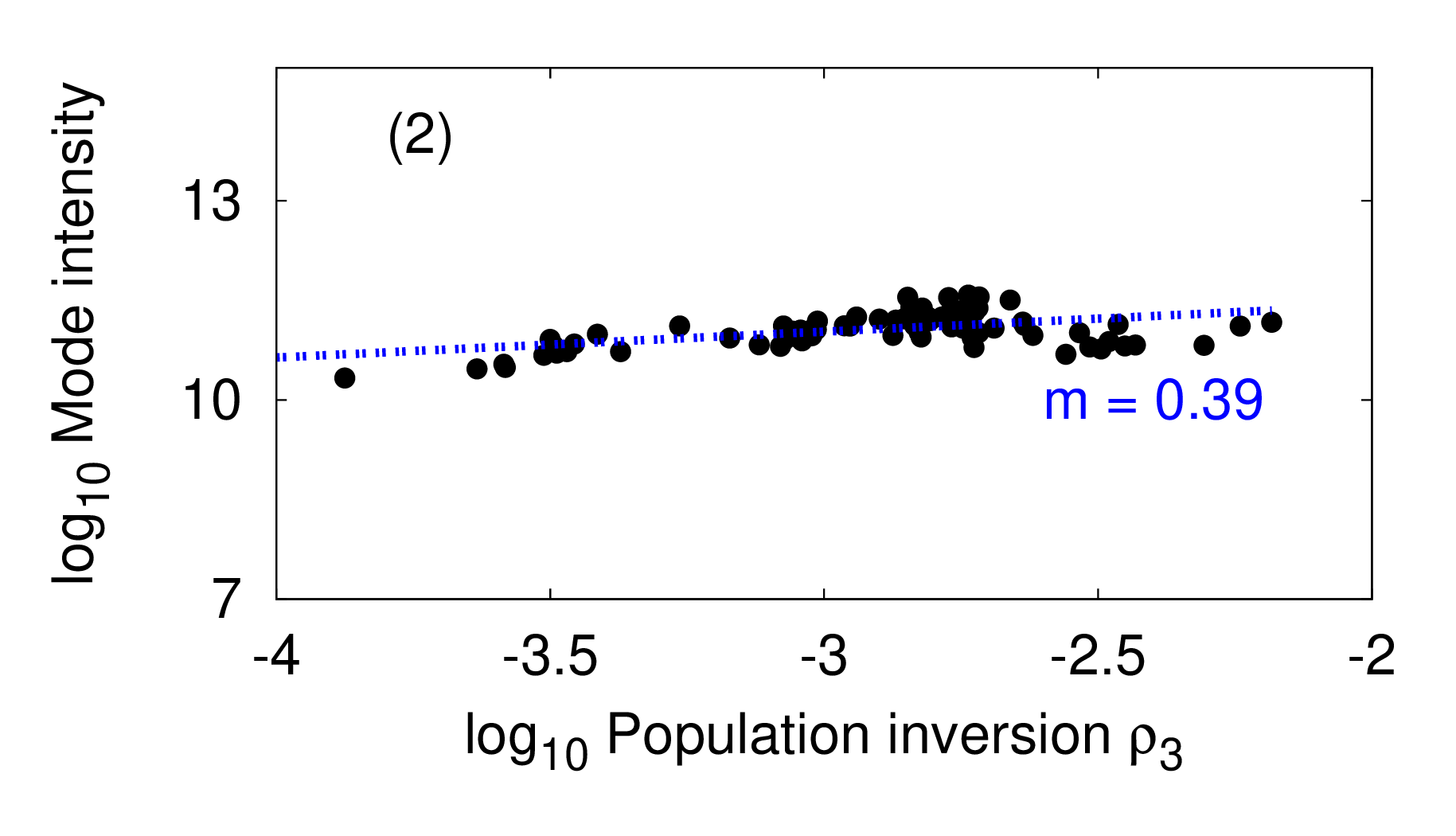}
    \includegraphics[width=4.5cm,angle=270]{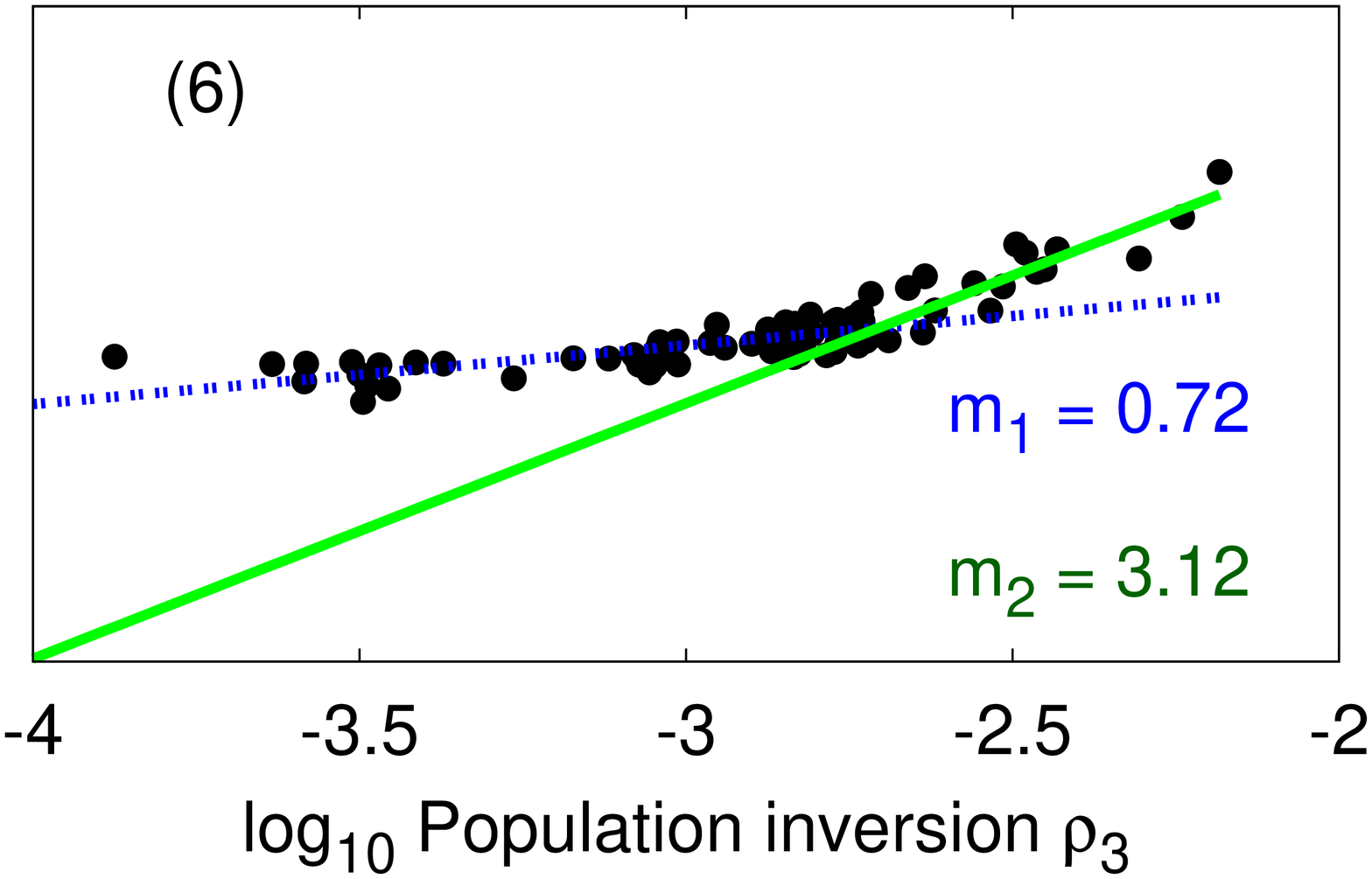}
    \caption{\label{fig:fig7}
      Mode intensities vs. population inversion 
      with noise for representative modes.
      Modes 1 and 3 are lasing without noise while modes 2 and 6 are suppressed without noise.
      Slopes of linear fits on a log-log scale indicate the power $m$ of intensity increase,
      i.e., intensity $\propto \rho_3^m$. 
      The lasing modes 1 and 3 experience superlinear increase above a threshold pump level.
      The lasing threshold is defined as the intercept of the two linear fits.
      Suppressed mode 2 does not but mode 6 does, indicating it is lasing when noise is included.
    }
  \end{center}
\end{figure}

The results in section \ref{ssc:spectbeh} illustrate that suppression of lasing modes due to gain saturation
is weakened in the presence of noise.
Some resonant modes, which fail to lase without noise, manage to lase
in the presence of noise, however, a proper definition of the lasing threshold is lacking.
The co-existence of multiple lasing modes and their interactions through the gain material make it difficult to
define the threshold for each separate mode using previously developed methods for single mode lasers
\cite{jin94,ricec,straufPRL06,ulrichPRL07,beveratos}.
The data in figure \ref{fig:fig7} clearly displays an abrupt change of slope for the mode intensity versus pump level.
This allows us to define a lasing threshold in the presence of noise and multiple lasing modes.

The mode intensities in figure \ref{fig:fig7} are plotted on a log-log scale in order to better examine the rate of increase.
The slope indicates the power $m$ of increase, $m < 1$ is sublinear and $m > 1$ is superlinear.
When the pump level exceeds a threshold, the mode intensity changes from a sublinear to a superlinear increase.
This reflects the onset of light amplification by stimulated emission into the mode.
We define this threshold as the lasing threshold for the mode.

For modes 1 and 3, the thresholds are $\left<\rho_3(x,t)\right>=9.7\times 10^{-3}$ and $12.4\times 10^{-3}$, respectively.
Noise has increased the absolute lasing thresholds.
However, relative to one another, the thresholds are closer together with noise.
Without noise, mode 6 is suppressed, in other words it does not lase.
With noise, the threshold is $\left<\rho_3(x,t)\right>=19.7\times 10^{-3}$.
and is much closer to the first lasing threshold.
Thus, noise reduces the difference in thresholds of different modes,
which makes the system behave more similar to a linear gain system (TM method).

Noise weakens the nonlinear effect of gain saturation.
Although mode 6 manages to lase with noise,
the other suppressed modes (2 and 7) do not lase even with noise included.
Mode 2 is shown in figure \ref{fig:fig7}.
No clear turn-on exists; its slope remains fairly constant and sublinear.
The same behavior occurs for mode 7 (not shown). 
The remaining cases of modes 4 and 5 do display a change of slope, but the superlinear increase of intensity is weak.
The range of superlinear increase, due to stronger mode competition at higher pump levels, is not enough to find a reasonable linear fit so their thresholds are not defined.

\section{Spatial Behavior\label{sec:spatial}}

Due to gain saturation, the mechanism through which mode competition and mode suppression occur is spatial hole burning.
In previous sections, spatial properties have been averaged out and only the spectral steady-state properties examined.
However, above threshold, spatial hole burning creates ``dead'' regions since there is no gain left for
larger-threshold lasing modes.
Next, we investigate the spatial properties of the population inversion and on it, the effects of noise.

\subsection{Well Above the Lasing Threshold}

\begin{figure}
  \begin{center}
    \includegraphics[width=5cm,angle=270]{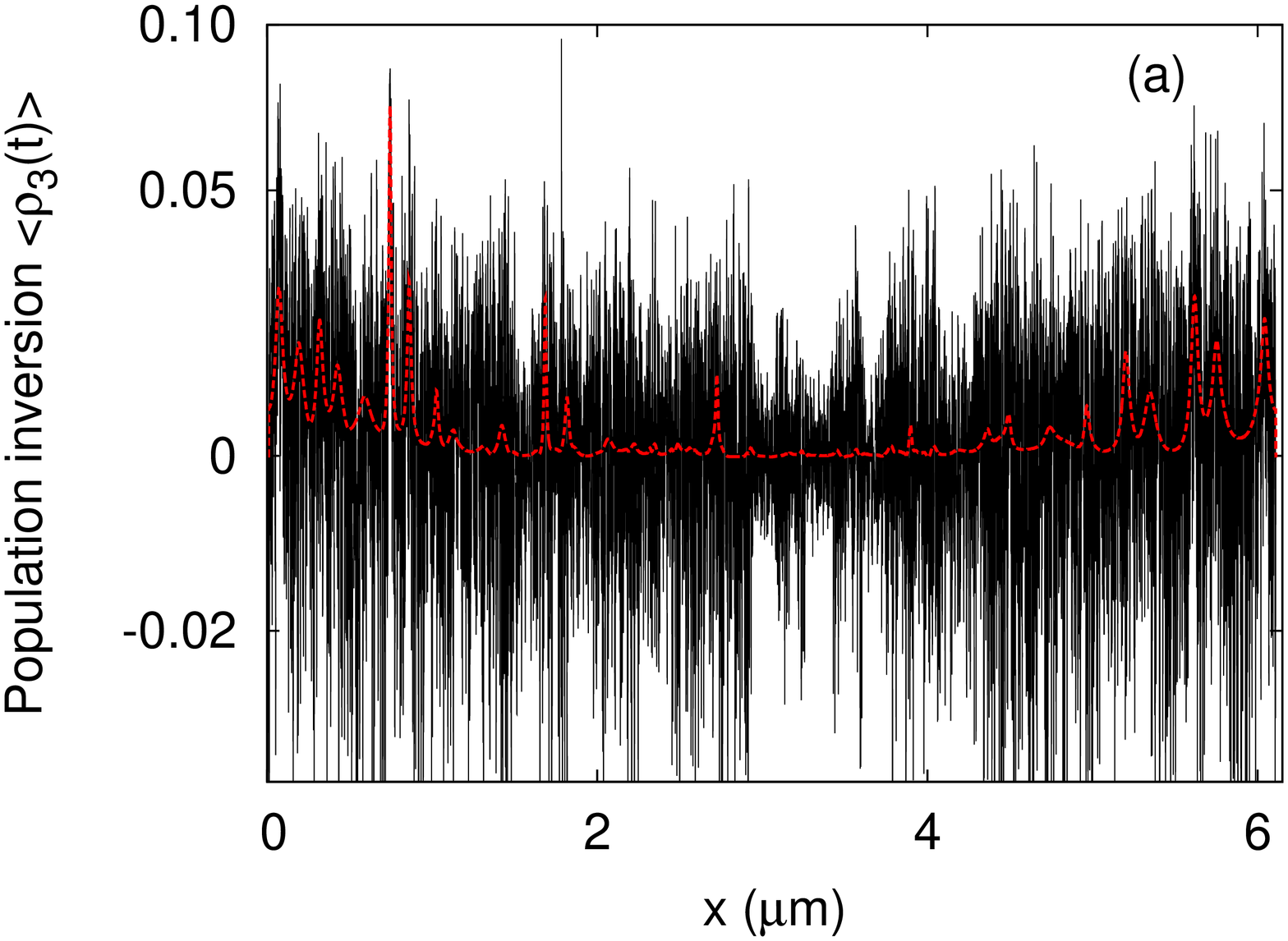}
    \includegraphics[width=5cm,angle=270]{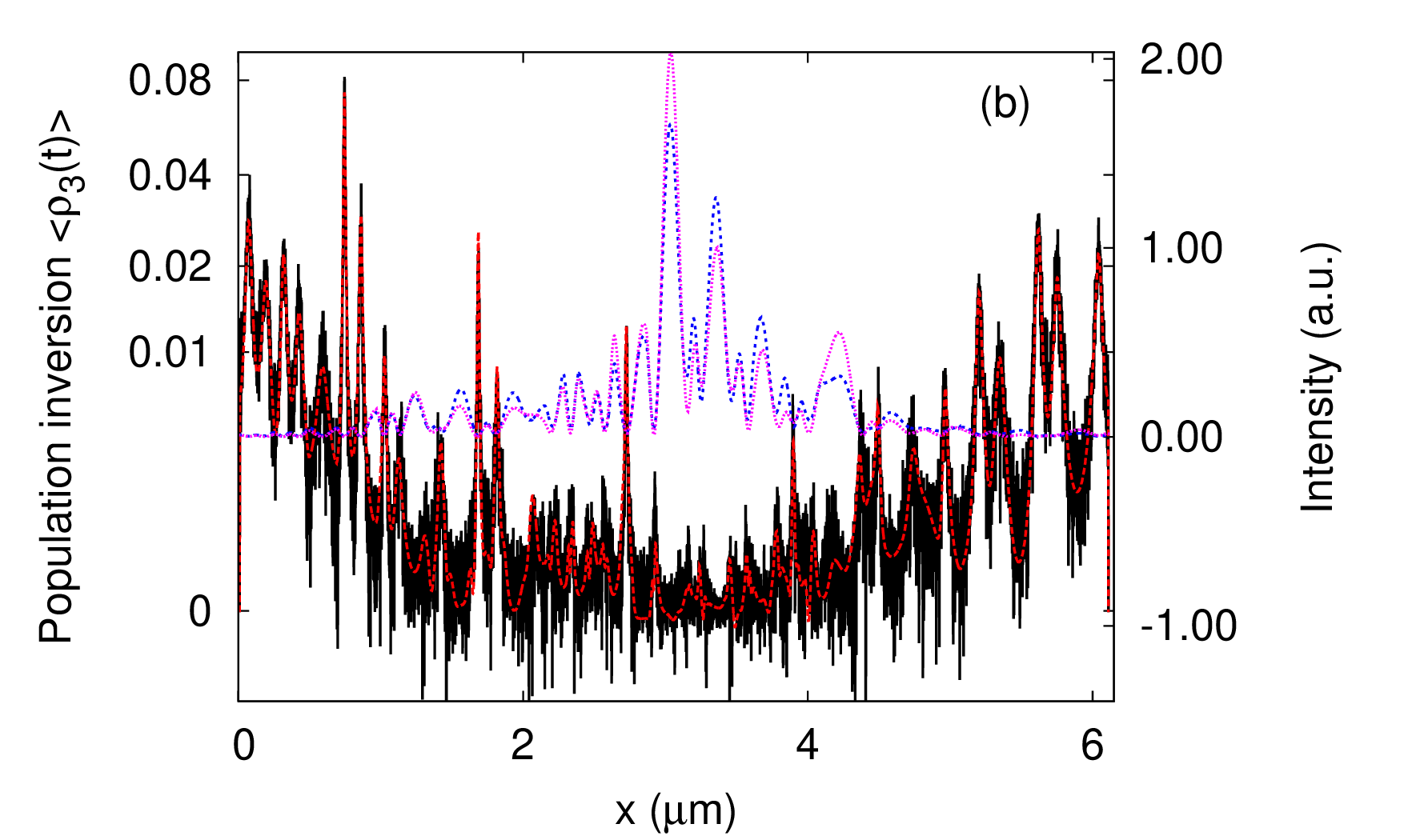}
    \caption{\label{fig:fig8}
      Time-averaged population inversion $\left<\rho_3(t)\right>$ vs. position $x$ 
      without noise (red) and with noise (black) for $\left<\rho_3(x,t)\right>=3.68\times 10^{-3}$.
      (a) $\rho_3$ with noise is averaged in time over one optical cycle $T$.
      (b) $\rho_3$ with noise is averaged in time over $2T_1$.
      Intensity without noise (magenta) and with noise (blue) is also shown;
      spatial hole burning is much weaker in (a).
    }
  \end{center}
\end{figure}

Our previous study demonstrated \cite{andreasenrln} that the spatial behavior of the population inversion
is similar with and without noise at high pump levels, well above the lasing threshold.
Only at low pump levels does ASE dominate the emission spectrum 
With an increasing pump level, gain saturation quickly sets in to suppress the fluctuations.
Without noise, the population inversion reaches a fairly stationary level and temporal averaging over
one optical cycle $T\approx \lambda_a/c$ gives an accurate assessment of inversion behavior.
With noise, averaging only over $T$ yields a more transient behavior of the gain medium
[see figure \ref{fig:fig8}(a)].
Much larger spatial fluctuations of the population inversion averaged over $T$ reflect stronger temporal
fluctuations on the time scale of $T$;
the inversion even becomes negative at some locations.
The optical cycle $T=2.0$ fs, dephasing time $T_2=1.9$ fs, and average cavity lifetime $\tau=17$ fs
are all similar. 
The atomic population changes over the much longer timescale of $T_1=1$ ps,
which is the longest time scale in the system.
Thus, with noise, the population inversion is averaged over $2T_1$ to remove short-time dynamic behavior.
Results are shown in figure \ref{fig:fig8}(b) along with the field intensity
to illustrate spatial hole burning.
When averaging over $2T_1$, the spatial behavior of gain is quite similar with and without noise.
This means noise does not remove the ``dead'' regions at high pump levels.
This is expected since the influence of noise decreases far above threshold
when the lasing signal is much larger than that of noise.

\subsection{Near the Lasing Threshold}

\begin{figure}
  \begin{center}
    \includegraphics[width=4.8cm,angle=270]{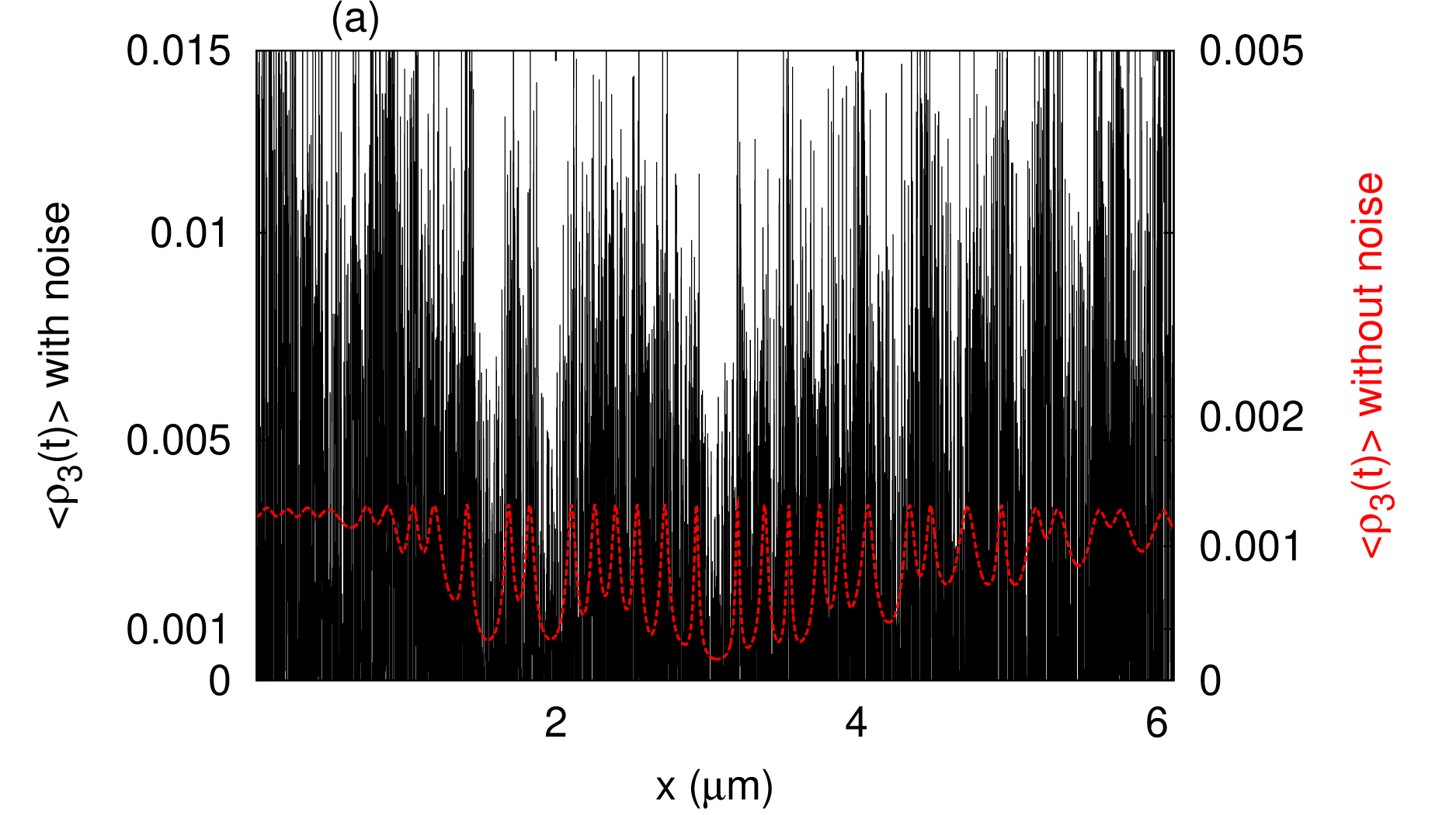}
    \includegraphics[width=4.8cm,angle=270]{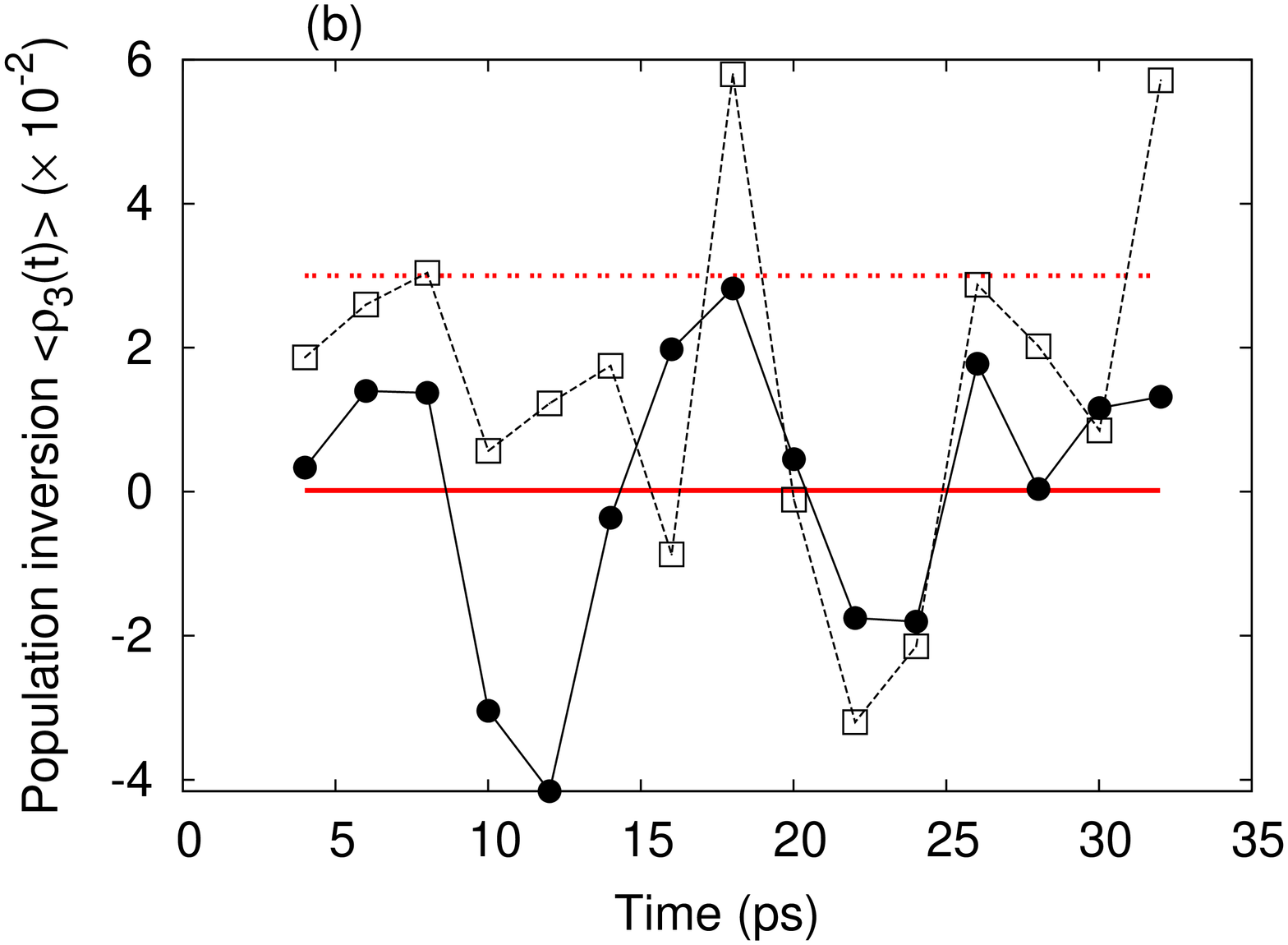}
    \caption{\label{fig:fig9}
      (a) Time-averaged population inversion $\left<\rho_3(t)\right>$ without noise (red) and
      with noise (black) for $\left<\rho_3(x,t)\right>= 0.86\times 10^{-3}$.
      (b) $\left<\rho_3(t)\right>$ with noise at $x=3$ $\mu$m (circles) and $x=0.5$ $\mu$m (squares).
      The solid horizontal line marks $\left<\rho_3(t)\right>=10^{-4}$ without noise at $x=3$ $\mu$m
      and the dotted horizontal  line marks $\left<\rho_3(t)\right>=0.03$ without noise at $x=0.5$ $\mu$m.
      $\rho_3$ is averaged over $2T_1=2$ ps with noise in (a) and (b). 
    }
  \end{center}
\end{figure}

The spatial behavior of the population inversion with and without noise for a low pump level
is shown in figure \ref{fig:fig9}(a).
Without noise, $\rho_3(x)$ is averaged over $T$ and with noise, it is averaged over $2T_1$.
Even averaging over the longest time scale in the system in this case, does not make $\rho_3$ 
with noise converge to that without noise.
Note that the population inversion without noise is plotted on a different scale for comparison,
since it is over an order of magnitude smaller.
The inversion with noise fluctuates dramatically in space with some spatial points (not shown) becoming negative
[similar to figure \ref{fig:fig8}(a)]. 

Not only does the inversion fluctuate strongly in space, but also in time.
In a spatial region not greatly influenced by spatial hole burning ($x=0.5$ $\mu$m),
the inversion in figure \ref{fig:fig9}(b) fluctuates greatly in time, even though averaged over $2T_1$.
These fluctuations still occur when spatial hole burning is strong, for example, at $x=3$ $\mu$m.
At this location, the inversion is dynamically ``dead'' without noise, i.e., $\rho_3 \sim 0$, due to the low-threshold
lasing modes depleting the gain.
Figure \ref{fig:fig9}(b) illustrates the changes of the inversion $\left< \rho_3(t)\right>_{2T_1}$ in time at $x=3$ $\mu$m due to noise.
The gain medium is constantly altered by spatial and temporal fluctuations and dead regions are overcome.
In other words, spatial hole burning is unable to continually enforce gain depletion in the presence of noise.

\subsection{Gradual behavioral change}

From the results above, it is clear that the spatial profile of the population inversion with noise is most different
from that without noise near the lasing threshold.
Without noise, the population inversion is depleted in regions of high laser intensity. 
With noise, the gain has a more uniform spatial behavior.
Thus, noise weakens gain depletion.
This helps to overcome the spatial regions of gain that would be depleted without noise.
To quantize the difference of population inversion with noise $(\rho_3(x))_n$ and without noise $(\rho_3(x))_w$, 
we take the difference between the two,
\begin{equation}
  K = \frac{\int_0^L |(\rho_3(x))_n - (\rho_3(x))_w| dx}{\int_0^L (\rho_3(x))_w dx}.\label{eq:KKaKb}
\end{equation}

\begin{figure}
  \begin{center}
    \includegraphics[width=6cm,angle=270]{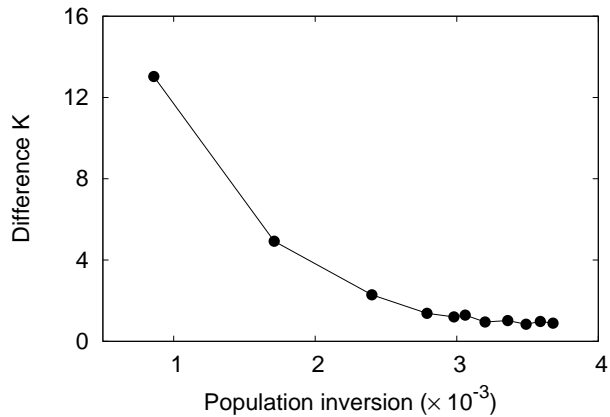}
    \caption{\label{fig:fig10}
      Spatial differences $K$ between the population inversion with and without noise calculated via equation (\ref{eq:KKaKb}).
      With spatial averaging, the two inversion distributions are roughly equal (horizontal axis). 
      Their spatial differences increase greatly at low pump levels.
    }
  \end{center}
\end{figure}

Figure \ref{fig:fig10} shows $K$ as a function of the pump level $\rho_3$.
The spatial distributions $(\rho_3(x))_n$ and $(\rho_3(x))_w$ are compared when their spatially \emph{averaged} quantities are roughly equal.
For low pump levels, the difference is greatest.
The distributions converge toward one another at high pump levels, as expected.

\section{Stochastic Resonance and Stochastic Linearization\label{sec:stoch}}

In nonlinear systems, the mechanism of stochastic resonance (SR) \cite{benziJPA81} manifests itself when noise
is able to amplify a weak signal past a certain threshold.
With noise included, it was observed in section \ref{sec:noise} that peaks appeared in the emission spectra which are absent from the spectra without noise.
The appearance of such peaks with noise suggests the mechanism of SR.
However, in typical SR cases, an external driving source forces a ``resonance'' peak in the power spectrum 
at the driving frequency if the noise is strong enough \cite{wiesenfeldNAT95,gammaitoniRMP98}.
The random lasers considered here employ an \emph{incoherent} pump to drive the system.
Thus, the typical mechanism of SR is not observed.
The peaks that appear in the emission spectrum are associated with intrinsic resonances of the underlying random structure \cite{review}.

Closely related to stochastic resonance is the mechanism of stochastic linearization.
Noise added to a continuous signal subject to nonlinear effects (in this case due to gain saturation) and a threshold condition (in this case the lasing threshold),
has the effect of linearizing the output \cite{gammaitoniPRE95}.
In digital signal processing, this is known as dithering \cite{bennetBST48} or stochastic linearization (SL).
In section \ref{sec:noise}, we observed that strong nonlinear effects, such as the suppression of lasing modes due to gain saturation,
are weakened with noise included in the calculation.
Moreover, in section \ref{sec:spatial}, it was observed that regions of depleted gain were overcome by noise,
a typical marker of SL.
In this section, we further explore the occurrence of SL in random lasers.

For a signal subject to a threshold condition, nonlinearity can cause unwanted errors in the detection of that signal.
On one hand, nonlinearity may push the original signal above threshold causing a false positive detection.
On the other hand, nonlinearity may pull the original signal below threshold causing a false negative.
Adding the proper amount of noise to such a system can remove the detection errors.
For example, with a signal originally above threshold, random noise can mitigate the effects of nonlinearity so that the 
signal is pulled below threshold less frequently.
Thus, if many measurements are taken, a positive detection of the signal being above threshold occurs most often.
A statistical average of measurements therefore yields the correct detection of the signal being above threshold.
Likewise, a signal originally below threshold can be correctly detected if the proper amount of noise is added to remove the effects of nonlinearity.
It is in this sense that a signal is linearized by a stochastic process since the effects of the nonlinearities are removed.

\begin{figure}
  \begin{center}
    \includegraphics[width=3.4cm,angle=270]{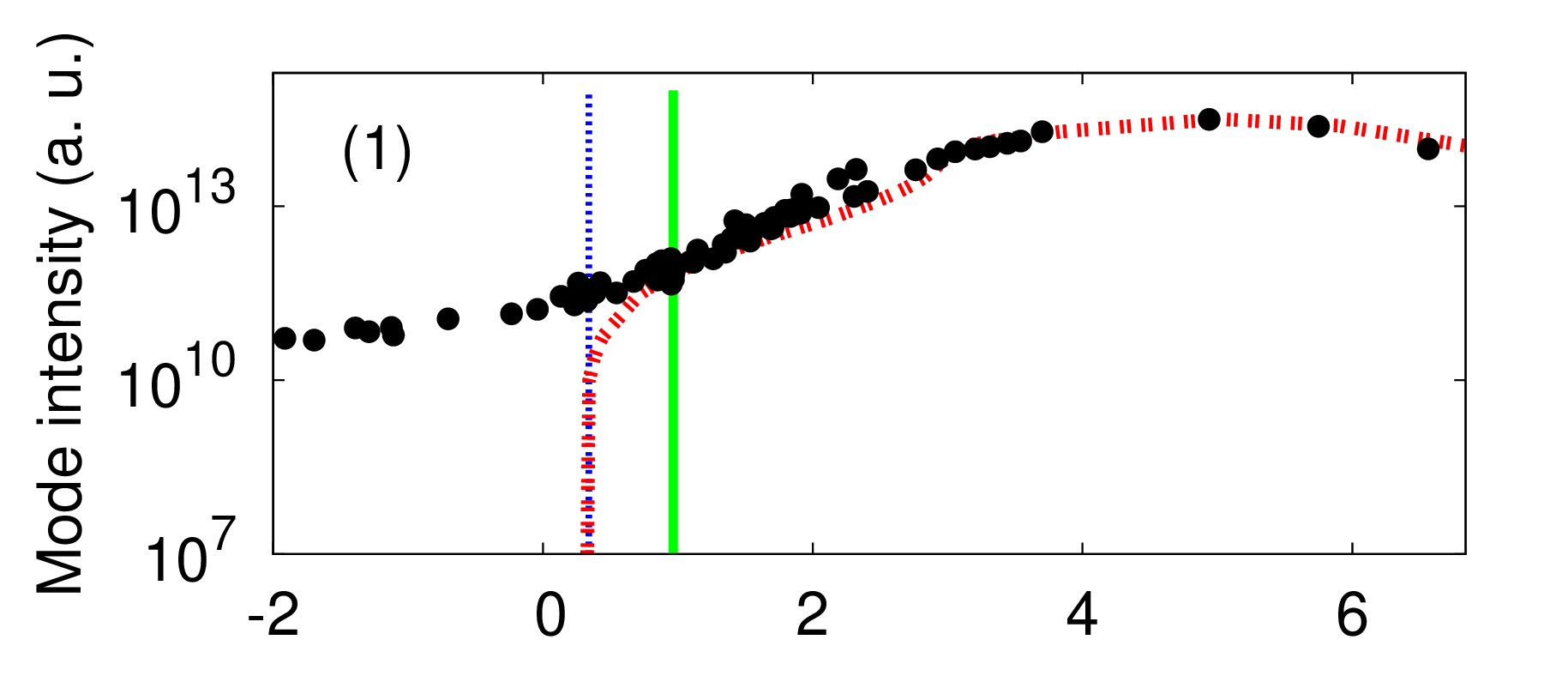}
    \includegraphics[width=3.4cm,angle=270]{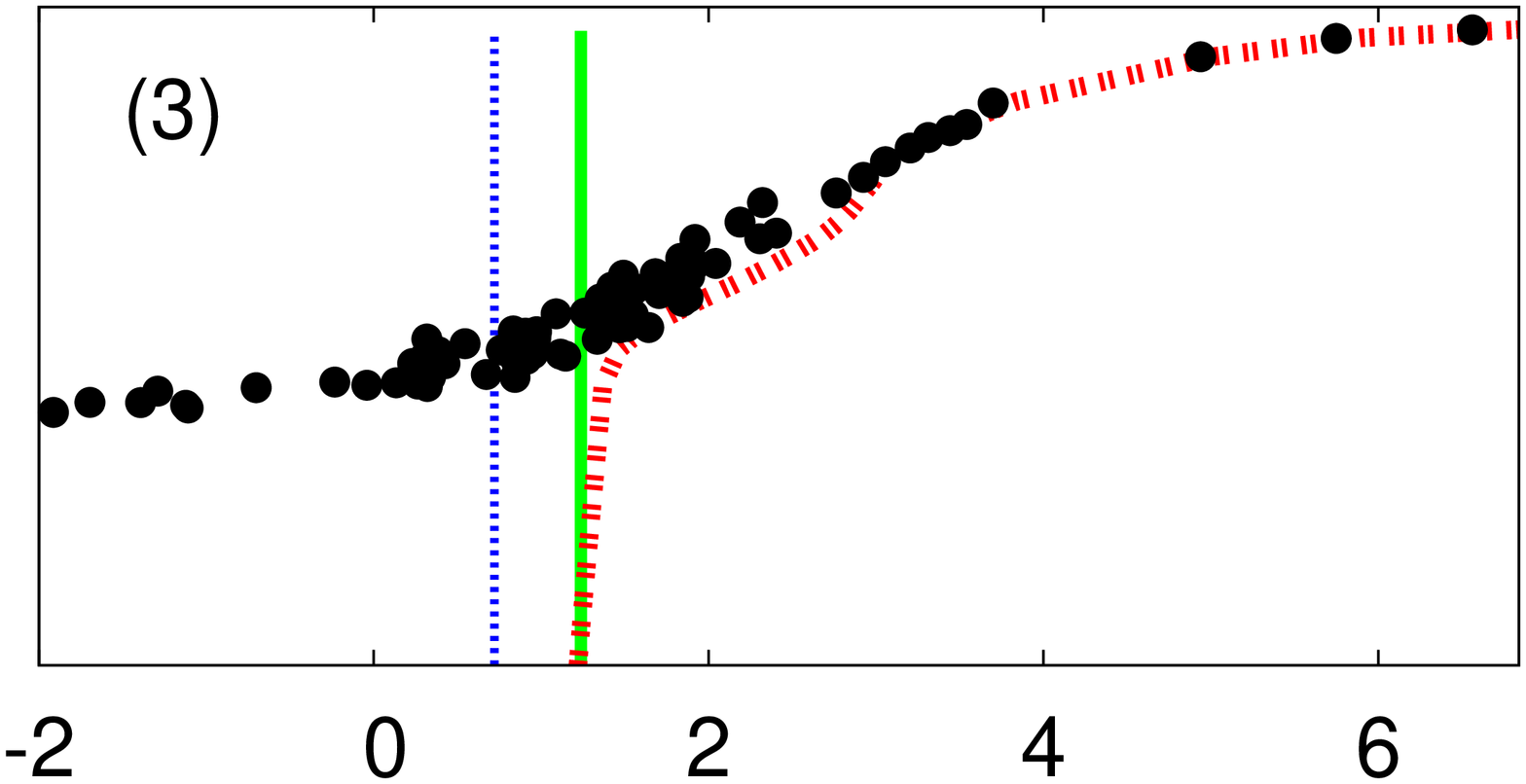}\\
    \includegraphics[width=4.5cm,angle=270]{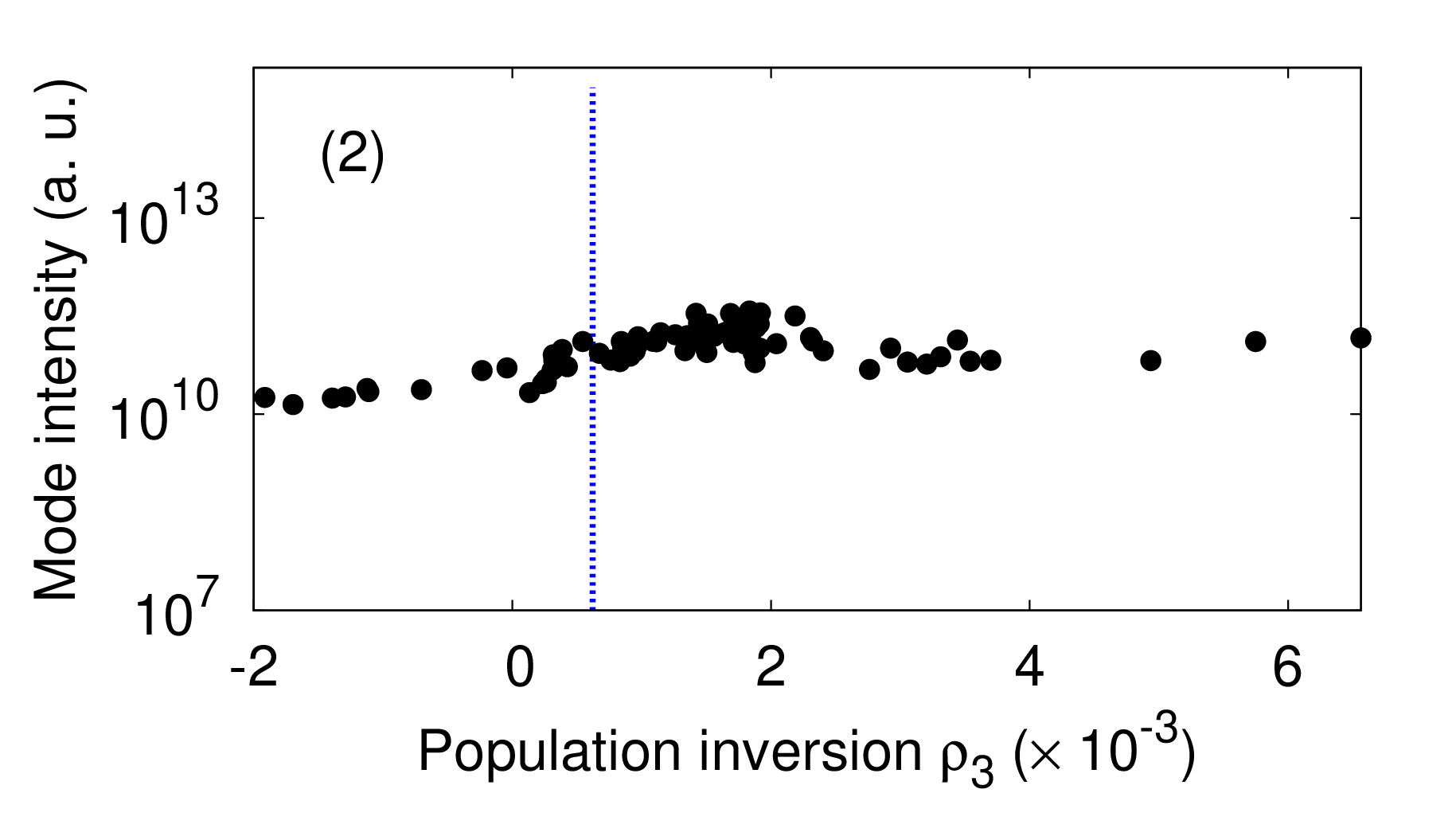}
    \includegraphics[width=4.5cm,angle=270]{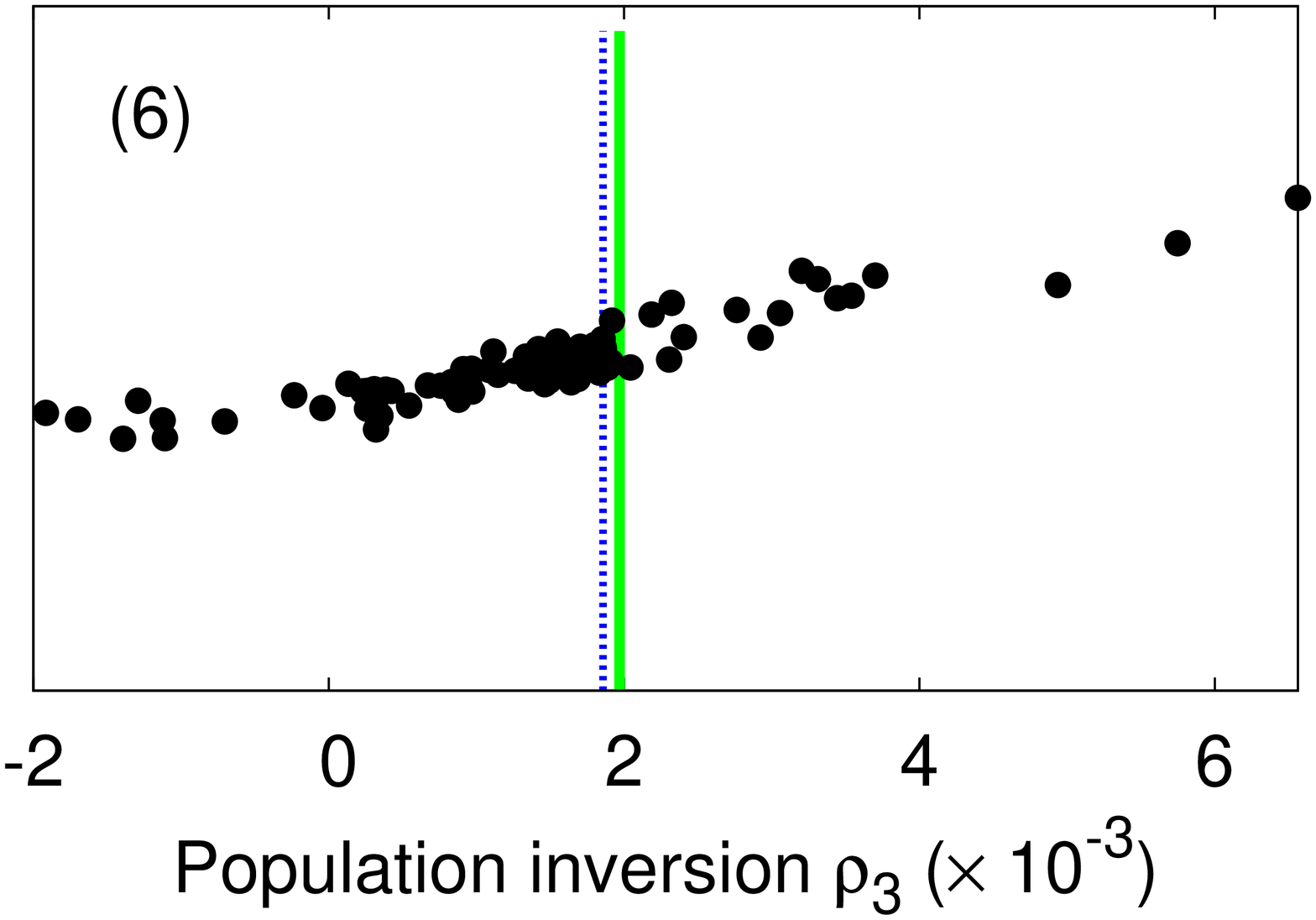}
    \caption{\label{fig:fig11}
      Mode intensities vs. population inversion 
      without noise (red dashes) and with noise (black solid circles) for lasing modes 1 and 3,
      and modes 2 and 6, which are suppressed without noise.
      The vertical blue dotted lines indicate the lasing thresholds without gain saturation (via the TM method).
      The vertical green solid lines indicate the lasing thresholds with noise (via the stochastic MB method).
      The nearly vertical increase of red dashes indicates the lasing threshold without noise (via the MB method).
    }
  \end{center}
\end{figure}

In random lasers, we check if noise linearizes the emission signal across the lasing threshold.
We define the lasing threshold to be reached as the threshold without gain saturation (found via the TM method).
Since we observed mode suppression, nonlinearity (gain saturation) in random lasers only influences the emission in one way, i.e., 
the signal is pulled below its threshold.
Therefore, as suggested by sections \ref{sec:noise} and \ref{sec:spatial}, 
we check if noise is able to push the signal up above threshold again thereby removing the influence of nonlinearity.

The effect of noise on representative random laser modes is shown in figure \ref{fig:fig11}.
Mode intensity with noise is clearly nonzero for all cases below the TM threshold (marked by the vertical blue line). 
A nonzero signal, however, does not indicate the threshold has been reached.
From the steady-state value of the population inversion, we see that
there is no sustained lasing oscillation in these cases since there is not enough gain to compensate the loss.
The noise can push the system above the lasing threshold for short periods of time, 
however, the population inversion that is time-averaged over $2T_1$ does not reach the threshold value for lasing.
This time-averaging is equivalent to the statistical averaging discussed above in relation to stochastic linearization.
The time-averaged nonzero signal below the lasing threshold is attributed to spontaneous emission for $\rho_3 < 0$ and ASE for $\rho_3 > 0$ \cite{andreasenrln}. 

The signals with noise in figure  \ref{fig:fig11} cross the TM lasing thresholds in a continuous manner, a behavior typical of SL.
However, the signals with noise for modes 1 and 3 have not yet reached their lasing threshold, as defined in section \ref{sec:noisethresh}
and indicated by vertical green lines.
For mode 1, gain saturation does not play a role since no other modes lase to suppress it \cite{andreasenNLE}. 
Noise influences mode 1 by pulling it below its threshold so that it does not lase.
This occurs because noise draws energy away from the lasing mode and distributes it over many other modes via ASE.
For mode 3, the effects of gain saturation and noise influence its behavior;
they both cause an increase of the lasing threshold.
However, the threshold with and without noise is nearly the same. 
Though the lasing threshold is increased with noise, the effects of gain saturation are mitigated with noise.
These two effects are balanced for this particular mode resulting in a similar threshold with and without noise.

Lasing oscillation in modes 2 and 6 in figure \ref{fig:fig11} are always suppressed without noise.
With noise, much larger signals appear in the emission spectra.
For mode 6, noise allows the lasing threshold to be reached when it is otherwise impossible (due to mode suppression).
Without noise, nonlinearity due to gain saturation caused an ``error,'' in that its signal was not detected when it otherwise would have been above the threshold without gain saturation.
Inherent noise weakens the effects of gain saturation enough so that the signal is detected.
In the case of mode 2, the proper amount of noise does not inherently exist in the system to remove this error.
The appearance of mode 2 with noise for $\rho_3 > 0$ is merely due to spontaneous emission,
since a superlinear behavior of the emission signal is never observed.

We believe the inability of noise to excite mode 2 to lase is because of improper ``tuning.''
The noise we consider in random lasers is inherent and therefore, not tuned to give optimal output.
This opens the question, however, if noise can be tuned, e.g.,
by adjusting the atomic interaction with the heatbath. 
With the proper amount of noise, mode 2 may lase and its amplitude maximized.

\section{Conclusion\label{sec:conclusion}}

Gain saturation causes strong nonlinear effects in random lasers in the multimode regime.
We have shown these effects, such as the increased lasing thresholds and mode suppression,
by comparing full-wave Maxwell-Bloch simulations to linear gain simulations that exclude gain saturation.
Inherent noise of the laser system was found to somewhat mitigate the nonlinear effects.
Noise increases the first lasing thresholds due to redirection of energy out of lasing modes, 
but reduces the thresholds of modes that lase at higher pump levels.
Noise constantly excites all modes and their dwell time in the random system results in peaks in the emission spectrum
that are absent without noise.
Above the transparency point, amplified spontaneous emission enhances the mode amplitude and allows a smooth transition to lasing.
In some cases, this process allows lasing of modes that are suppressed when noise is not included.
The result is inherent stochastic linearization.
We have shown that this is made possible when noise overcomes ``dead'' regions of gain caused by spatial hole burning.
We further suggest that noise may be tuned by adjusting the atomic medium providing gain, 
to possibly excite and maximize the amplitude of all possible lasing modes.
It may also be possible to frustrate lasing in particular modes by properly adjusting the noise.

\section*{Acknowledgments}
We thank Prof. C. Vanneste for stimulating discussions.
This work is supported in part by the National Science Foundation
under the Grant DMR-0808937,
and the Yale Biomedical HPC Center and NIH Grant RR19895.
JA acknowledges support from the Embassy of France in the United States.

\section*{References}
\bibliographystyle{unsrt}

\end{document}